\documentclass[useAMS,usenatbib]{mn2e}

\usepackage{amsmath,array,amsfonts}
\usepackage{amssymb}
\usepackage{graphicx}
\usepackage{placeins}
\usepackage{float}
\usepackage{natbib}
\usepackage{pstricks}
\usepackage{verbatim}
\usepackage{subfigure}
\usepackage{booktabs}
\usepackage{siunitx,booktabs,multirow, dcolumn}

\newcommand{\ltsima} {$\; \buildrel < \over \sim \;$}
\newcommand{\gtsima} {$\; \buildrel > \over \sim \;$}
\newcommand{\lta} {\lower.5ex\hbox{\ltsima}}
\newcommand{\gta} {\lower.5ex\hbox{\gtsima}}

\newcolumntype{d}[1]{D{.}{.}{#1} }

\def\f{\frac}
\def\nn{\nonumber}
\def\ln{\mathrm{ln}}

\title{Weak Lensing with Sizes, Magnitudes and Shapes}

\author[J. Alsing, D. Kirk, A. Heavens, A. Jaffe]{Justin Alsing$^1$\thanks{e-mail:  j.alsing12@imperial.ac.uk}, Donnacha Kirk$^2$, Alan Heavens$^1$, Andrew H. Jaffe$^1$ \\
$^1$     Imperial Centre for Inference and Cosmology, Department of Physics, Imperial College, \\Blackett Laboratory, Prince Consort Road, London SW7 2AZ, U.K.\\
$^2$     Department of Physics \& Astronomy, University College London, Gower Street, London, WC1E 6BT, UK\\}
\date{Accepted ;  Received ; in original form }

\begin{document}
\maketitle

\begin{abstract}
Weak lensing can be observed through a number of effects on the images of distant galaxies; their shapes are sheared, sizes and fluxes (magnitudes) are magnified and positions on the sky are modified by the lensing field. Galaxy shapes probe the shear field whilst size, magnitude and number density probe the convergence field. Both contain cosmological information. In this paper we are concerned with the magnification of sizes and magnitudes of individual galaxies as a probe of cosmic convergence. We develop a Bayesian approach for inferring the convergence field from measured sizes, magnitudes and redshifts and demonstrate that this inference requires detailed knowledge of the joint distribution of intrinsic sizes and magnitudes. We build a simple parameterised model for the size-magnitude distribution and estimate this distribution for CFHTLenS galaxies. In light of the measured distribution, we show that the typical dispersion on convergence estimation is $\sim0.8$, compared to $\sim 0.38$ for shear. We discuss the possibility of physical systematics for magnification (similar to intrinsic alignments for shear) and compute the expected gains in the Dark Energy Figure-of-Merit (FoM) from combining magnification with shear for different scenarios regarding systematics: accounting for intrinsic alignments but no systematics for magnification, including magnification could improve the FoM by upto a factor of $\sim2.5$, whilst when accounting for physical systematics in both shear and magnification we anticipate a gain between $\sim25\%$ and $\sim65\%$. The fact that shear and magnification are subject to different systematics makes magnification an attractive complement to any cosmic shear analysis.

\end{abstract}
\section{Introduction}
\label{intro}
Light from distant galaxies is continuously deflected by the gravitational potential of intervening large-scale density inhomogeneities in the Universe on its way to Earth. This weak gravitational lensing results in a coherent distortion of observed galaxy images on the sky, providing us with a powerful probe of the growth rate of potential perturbations and the geometry of the universe through the distance-redshift relation. Traditionally, and for good reasons, the statistic of choice for weak lensing has been cosmic shear - the distortion of observed shapes of source images. However, weak lensing has a number of other effects; principally, the sizes and fluxes of individual objects are magnified and the observed positions of galaxies on the sky are modified due to lensing. In an ideal analysis one would like to use all of the available information to elicit the full statistical potential from a weak lensing survey. Further motivation for developing multiple independent probes of weak lensing comes in the context of systematic effects. Systematic uncertainties pose a major challenge for any cosmic shear analysis. At the shape measurement level, accounting for the point-spread function, noise-rectification, seeing/optical distortions and selection effects can all introduce systematic errors \citep{Kaiser2000, Erben2001, Bernstein2002, VanWaerbeke2003, Hirata2003}. There are physical systematics too in the form of intrinsic alignments; nearby galaxies form in a similar tidal gravitational field, which may lead to a preferred ellipticity orientation of neighbouring galaxies \citep{Heavens2000, Catelan2001, Crittenden2001, Croft:2000gr}. This intrinsic alignment of galaxy ellipticities introduces a spurious signal when trying to extract a cosmic shear signal from the correlation of galaxy shapes. Different probes of weak lensing are likely to be subject to different systematics, providing strong motivation for developing several independent methods for extracting a weak lensing signal.
 
Here we are concerned with using the magnification of the sizes and magnitudes of individual sources to measure a cosmological lensing signal. This relatively new approach and has received some attention both theoretically \citep{Heavens:2013vq, Casaponsa2012} and observationally \citep{Schmidt:2011ju}, after an initial early study \citep{Bartelmann1996}. \citet{Casaponsa2012} showed that the convergence field can be recovered from the measured sizes of simulated galaxy images without any evidence of bias, provided the galaxies are larger than the point-spread function (PSF) and have S/N larger than 10. These are very similar requirements for accurate estimation of shear, and they point out that since the shape measurement process also inevitably requires investigation of the size, the size information comes with little additional effort. \citet{Heavens:2013vq} (hereafter HAJ13) developed the statistics of a combined shear-size-magnification analysis and demonstrated that substantial gains in the Dark Energy Figure-of-Merit may be expected from combining size and shape information, subject to assumptions about the intrinsic scatter of galaxy sizes. We also showed that for galaxies with an exponential brightness profile, size and shape estimates should be approximately uncorrelated, so one could anticipate the full benefit from combining size and shape information. \citet{Schmidt:2011ju} measured a galaxy-galaxy lensing signal using the magnification of sizes and magnitudes in COSMOS galaxies, finding a signal consistent with shear but with roughly 40\% of the signal-to-noise. This paper attempts to address three key questions associated with cosmic magnification. (1) Given a size, magnitude and redshift, how do we estimate the convergence field $\kappa$? (2) What is the intrinsic distribution of sizes and magnitudes for galaxies in a typical lensing survey and how does the shape of this distribution impact our ability to recover $\kappa$ from size and magnitude measurements? (3) What statistical gains might be expected from combining magnification with cosmic shear? Cosmic magnification may be subject to physical systematic effects similar to intrinsic alignments for shear, in the form of size- or magnitude-density correlations (ISCs). We explore the extent to which both IAs and ISCs may impact the constraining power of a combined shear-magnification analysis for different levels of systematic uncertainty.
 
The organization of this paper is as follows. In \S \ref{probes} we briefly introduce the ideas of galaxy shape information as a probe of cosmic shear and size/flux magnification as a probe of cosmic convergence. In \S \ref{magnification_estimation} we develop a Bayesian approach for estimating the convergence field from an observed size, magnitude and redshift, and contrast this with a simple unbiased estimator. In \S \ref{joint_dist} we measure the intrinsic size-magnitude distribution of CFHTLenS galaxies and in \S \ref{dispersion} we explore how well we would expect to be able to recover $\kappa$ from sizes and magnitudes drawn from this intrinsic distribution. In \S \ref{fisher} we extend the work of HAJ13 to include the new information about the size-magnitude distribution of CFHTLenS galaxies, and forecast the expected information gain from including size-magnitude magnification with cosmic shear for various scenarios regarding intrinsic alignments and size- and magnitude-density correlations.

\section{Shapes, sizes and magnitudes as probes of cosmic shear and convergence}
\label{probes}
Gravitational lensing effects can be described by the Jacobian matrix mapping source angular positions $\boldsymbol{\theta}^\mathrm{S}$ to image positions $\boldsymbol{\theta}^\mathrm{I}$, i.e., $d\theta_i^\mathrm{S}=\mathcal{A}_{ij}d\theta_j^\mathrm{I}$. In the weak lensing limit, this distortion matrix can be decomposed as
\begin{align}
\mathcal A=\left( \begin{array}{c c}
1-\kappa-\gamma_1 & -\gamma_2 \\
-\gamma_2 & 1-\kappa+\gamma_1\\
\end{array} \right),
\end{align}
which defines the convergence field $\kappa$ and complex shear field $\gamma\equiv \gamma_1+i\gamma_2$. The magnification of surface area elements, $\mu$, is given by the determinant of this matrix,
\begin{equation}
\mu = \frac{1}{\det(\mathcal A)}=[(1-\kappa)^2-|\gamma|^2]^{-1},\label{mag}
\end{equation}
which in the weak lensing limit $|\kappa|,\,|\gamma|\ll 1$ (assumed throughout) can be approximated by $\mu\simeq 1+2\kappa$. This magnification of image area is accompanied by an increase in flux, since surface brightness must be conserved under lensing. The lensed area and luminosity of a source will hence be scaled by a factor $(1+2\kappa)$ under lensing, i.e., $A\rightarrow A(1+2\kappa)$ and $L\rightarrow L(1+2\kappa)$. Galaxy shapes, defined by their complex ellipticity $\epsilon$, will be `sheared' under lensing and in the weak lensing limit are linearly shifted by the complex shear $\gamma$, i.e., $\epsilon\rightarrow\epsilon+\gamma$. 

HAJ13 showed that for galaxies with exponential brightness profiles, joint estimation of the square-root of the image area and ellipticity are uncorrelated, making $\sqrt{\mathrm{Area}}$ an ideal measure of `size'. Throughout this paper we will denote `log-size' by $\lambda = \ln\sqrt{\mathrm{Area}}$, and will work with magnitudes rather than fluxes for convenience. The impact of lensing on an individual source in the weak lensing regime can then be summarized as:
\begin{align}
\label{weak_lensing_model}
\epsilon\rightarrow\epsilon + \gamma \nn \\
\lambda \rightarrow\lambda + \kappa \nn \\
m \rightarrow m -q\kappa
\end{align}
where $q = -5\;\mathrm{log}_{10}(e)\approx-2.17$ (converting from flux to magnitude).
Here we are concerned with how much information may be available from using both the magnitudes and the sizes of individual sources.

The magnification of source fluxes can be exploited in two ways. In a flux-limited survey, positive magnification in a patch of sky will push sources across the flux limit, increasing the observed number density of sources in that patch. This will be accompanied by a demagnification of source number density due to the divergence in source positions due to lensing. Provided these competing effects do not cancel each other precisely, observations of the local number density of sources in a flux limited survey can be used to extract information about the lensing magnification field. This approach has become known as simply as \emph{flux magnification} and has received attention both theoretically and observationally \citep{Hildebrandt2009, Hildebrandt2013, VanWaerbeke2010, Duncan:2013bb}. This altering of the clustering of observed sources is \emph{not} the subject of this paper, but could in principle provide further information.

Alternatively, one can use the magnification of fluxes of individual sources to extract information about the lensing field. This, along with the magnification of image sizes, is the focus of this paper and will henceforth simply be referred to as magnification.
 
\section{Estimating convergence from sizes and magnitudes}
\label{magnification_estimation}
In order to perform a weak lensing analysis using galaxy sizes and magnitudes, we are faced with the following problem: given a sample of galaxies with measured sizes, magnitudes and redshifts, we want to estimate the lensing field across the sky. The intrinsic shape, size or flux of a given observed galaxy is not known a priori, so the lensing effect on individual sources is not accessible. However, if the intrinsic distribution of these properties is known, then a weak lensing analysis can be performed at a statistical level by observing a large number of galaxies and looking for a signal through deviations in the observed distribution. In the following section we develop a Bayesian framework for estimating $\kappa$ from a measured size, magnitude and redshift of an individual source. For comparison we also develop a simple unbiased (but sub-optimal) estimator combining size and magnitude information.
\subsection{Bayesian convergence inference}
\label{bayesian}
Given a size, magnitude and redshift for an individual object, we would like to write down a posterior distribution for $\kappa$, i.e., (using Bayes' theorem)
\begin{align}
p(\kappa|m, \lambda, z) = \frac{p(m, \lambda, z|\kappa)p(\kappa)}{p(m, \lambda, z)}.	
\end{align}
Principally, in order to write down the posterior on $\kappa$ from a single object we must know its observed (lensed) size and magnitude and redshift, the intrinsic (joint) distribution of the unlensed sizes, magnitudes and redshifts, and we must have a model (Eq. \eqref{weak_lensing_model}) for how lensing affects the observed size and magnitude. Let us begin with a simple set-up and build up the complexity step by step. Suppose we have a sample of galaxies with measured sizes and magnitudes, selected to be in a single redshift bin, from a survey with hard size and magnitude cuts. We neglect the small errors on the measured sizes and magnitudes for the time being, and since we are taking a single redshift bin we will drop $z$ to begin with (note this extends to tomography easily by considering a number of distinct redshift bins). In this case we are looking for
\begin{align}
p(\kappa|m, \lambda) = \frac{p(m, \lambda|\kappa)p(\kappa)}{p(m, \lambda)}.	
\end{align}
The likelihood of observing a source with a particular size and magnitude $(m, \lambda)$ given some value of $\kappa$ is essentially the probability of that source having an intrinsic (unlensed) size and magnitude $(m+q\kappa, \lambda - \kappa)$. However, we must also account for the fact that lensing will push sources across the magnitude and size cuts, modifying the normalization of the size-magnitude distribution as the overall number of sources will change as sources are pushed across the cuts. The likelihood is the intrinsic size-magnitude distribution, translated by $(q\kappa,- \kappa)$ and renormalized to one over the box bounded by the hard cuts $m_\mathrm{min}$, $m_\mathrm{max}$, $\lambda_\mathrm{min}$ and $\lambda_\mathrm{max}$. We can write the likelihood, in the limit of negligible measurement error, as:
\begin{align}
\label{likelihood_hard_cuts}
p(m, \lambda|\kappa) = \frac{p(m+q\kappa, \lambda - \kappa)}{\int_{m_\mathrm{min}}^{m_\mathrm{max}}\int_{\lambda_\mathrm{min}}^{\lambda_\mathrm{max}} p(m'+q\kappa, \lambda' - \kappa)dm'd\lambda'},
\end{align}
and the posterior for $\kappa$ given a size and magnitude is hence
\begin{align}
p(\kappa|m, \lambda) \propto \frac{p(m+q\kappa, \lambda - \kappa)p(\kappa)}{\int_{m_\mathrm{min}}^{m_\mathrm{max}}\int_{\lambda_\mathrm{min}}^{\lambda_\mathrm{max}} p(m'+q\kappa, \lambda' - \kappa)dm'd\lambda'}.	
\end{align}
Recall that this is the posterior distribution of $\kappa$ given an observed size and magnitude (neglecting errors in those quantities), for sources selected to be in a single redshift bin and from a complete sample down to hard size and magnitude cuts. For the purposes of forecasting how well we would expect to be able to estimate $\kappa$, it should be sufficient to work within this set of simplifying assumptions. In the remainder of this section we will lift these assumptions and develop a more general formalism for estimating $\kappa$ from sizes, magnitudes and redshifts.

\subsubsection{Smooth selection function}
In practice, we would not have hard cuts but rather a smooth selection function, $S(m, \lambda)$, which we will assume to be invariant under lensing. In this case the likelihood is the product of the shifted intrinsic distribution and the selection function, again renormalized to one to account for sources being shifted into the sample under the selection function:
\begin{align}
\label{likelihood_smooth_selection}
p(m, \lambda|\kappa) = \frac{p(m+q\kappa, \lambda - \kappa)S(m, \lambda)}{\int_{\mathbb{R}^2} p(m'+q\kappa, \lambda' - \kappa)S(m', \lambda')dm'd\lambda'},
\end{align}
Note that this reduces to \eqref{likelihood_hard_cuts} if we replace the selection function with a 2D tophat (i.e., in the limit of hard cuts). A smooth selection function that varies with $\kappa$ can straightforwardly be included in this formalism provided one has a model for how $S(m, \lambda)$ is modified under lensing. This is likely to be a small effect and we neglect it here.
\subsubsection{Including redshift information}
The intrinsic size-magnitude distribution will in general be a function of redshift and one would ideally like to use full redshift information when estimating $\kappa$. In this case the problem is extended to writing down a posterior on $\kappa$ given a size, magnitude and redshift: we must hence know the joint size-magnitude-redshift distribution, and the likelihood is extended to
\begin{align}
\label{likelihood_smooth_selection_redshift}
p(m, \lambda, z|\kappa) = \frac{p(m+q\kappa, \lambda - \kappa, z)S(m, \lambda, z)}{\int_{\mathbb{R}^3}p(m'+q\kappa, \lambda' - \kappa, z')S(m', \lambda', z')dm'd\lambda' dz'}.
\end{align}
\subsubsection{Including uncertainties in size, magnitude and redshift}
For a typical photometric redshift survey, the redshifts will have considerable uncertainties. The measured sizes and magnitudes will also be subject to error. We can include uncertainties on the measured quantities by extending our posterior to estimate the `true' redshift, magnitude, size and $\kappa$ simultaneously from noisy measurements $\hat{m}, \hat{\lambda}, \hat{z}$, and marginalize over the true size, magnitude and redshift since in the end we are only interested in $\kappa$. 
We can write
\begin{align}
\label{posterior_smooth_selection_redshift_uncertainties}
p(m, \lambda, z,\kappa|\hat{m}, \hat{\lambda}, \hat{z}) = \frac{p(\hat{m}, \hat{\lambda}, \hat{z}|m, \lambda, z)p(m, \lambda, z, \kappa)}{p(\hat{m}, \hat{\lambda}, \hat{z})}.
\end{align}
Note the assumed conditional independence of $\hat{m}, \hat{\lambda}$ and $\hat{z}$ on $\kappa$ once $m$, $\lambda$ and $z$ are fixed. We can also re-write the `prior' as $p(m, \lambda, z, \kappa) = p(m, \lambda, z|\kappa)p(\kappa)$ where the first term on the right-hand side is simply the `likelihood' from Eq. \eqref{likelihood_smooth_selection_redshift}. With these modifications, and marginalizing over $m, \lambda$ and $z$ we obtain
\begin{align}
\label{likelihood_smooth_selection_redshift_uncertainties_marginal}
&p(\kappa|\hat{m}, \hat{\lambda}, \hat{z}) = \nn \\
&\int\frac{p(\hat{m}, \hat{\lambda}, \hat{z}|m, \lambda, z)p(m+q\kappa, \lambda - \kappa, z)S(m, \lambda, z)}{\int p(m'+q\kappa, \lambda' - \kappa, z')S(m', \lambda', z')dm'd\lambda' dz'}p(\kappa)dmd\lambda dz.
\end{align}
There are a couple of important things to note about this result. Firstly, the posterior distribution depends on the intrinsic distribution of sizes and magnitudes. If this is obtained empirically, then issues associated with physics such as dust modifying the observed galaxy sizes and magnitudes will all be implicitly included; provided we understand how the observed quantities respond to lensing, comparing the observations of some quantity to its intrinsic distribution will elicit information about the lensing field regardless of what those measured quantities correspond to physically. However, if there are physical processes (such as the presence of dust halos) which impact observed galaxy sizes/magnitudes and those processes are related to the density field (for example if dust traces matter), they could introduce a systematic signal through intrinsic size-density or magnitude-density correlations. Secondly, note that in principle we need information about the size-magnitude distribution well into the incomplete regime (or below the size and magnitude cuts in the case of a hard cuts). This requires us to make assumptions about how the intrinsic distribution behaves below the reliable extent of our data. A deeper, higher-resolution survey could be effectively used to provide the necessary data. Furthermore, if we want to use all sources for which we have a measured size and magnitude, the selection function in the size-magnitude plane must be known in detail.
\subsection{Constructing a simple unbiased estimator for convergence}
\label{estimator}
In principle one would like to use the full posterior information on $\kappa$ for each individual source in the cosmological parameter inference process. In the case of cosmic shear it is commonplace to use the measured ellipticity of each source as a point estimator for the shear field. For comparison, we construct a similar point estimator for the convergence field using measured sizes and magnitudes. 

In the weak lensing limit, the expected sizes and magnitudes of lensed sources (above some flux and size cuts) are simply linearly shifted by the convergence:
\begin{align}
\label{mean_shifts}
&\langle\lambda\rangle \rightarrow \langle\lambda\rangle + \eta_\lambda\kappa \nn \\
&\langle m\rangle \rightarrow \langle m\rangle - \eta_m\kappa.
\end{align}
where the `responsivities' $\eta_\lambda = \partial\langle\lambda\rangle/\partial\kappa$ and $\eta_m=\partial\langle m\rangle /\partial\kappa$ account for both the boosting of size/magnitude due to lensing and the effect of smaller/fainter sources being pushed over the flux and size limits. The averages $\langle\lambda\rangle$ and $\langle m\rangle$ are taken over the observed distribution. In the absence of size and magnitude limits $\eta_\lambda = 1$ and $\eta_m = 2.17$. From \eqref{mean_shifts} we could write down two simple unbiased estimators for the convergence:
\begin{align}
\label{two_estimators}
\hat{\kappa} = (\lambda - \langle\lambda\rangle)\eta_\lambda^{-1}\nn \\
\hat{\kappa} = (m-\langle m\rangle)\eta_m^{-1}.
\end{align}
It is then straightforward to construct a single estimator taking a linear combination of these two:
\begin{align}
\label{linear_combination}
\hat{\kappa} = \alpha_\lambda(\lambda - \langle\lambda\rangle)\eta_\lambda^{-1} + \alpha_m(m-\langle m\rangle)\eta_m^{-1}
\end{align}
where $\alpha_\lambda$ and $\alpha_m$ are some weights to be determined. Note that $\langle\lambda\rangle$, $\langle m\rangle$, $\eta_\lambda$, $\eta_m$, $\alpha_\lambda$ and $\alpha_m$ may all be functions of redshift, so in general one should bin sources in redshift. 
We choose weights $\alpha_\lambda$ and $\alpha_m$ that minimize the variance of the estimator; minimizing the variance of \eqref{linear_combination} subject to the constraint $\alpha_\lambda + \alpha_m = 1$ to ensure the estimator is unbiased with respect to $\kappa$, we obtain weights
\begin{align}
\label{weights}
\alpha_\lambda &= \frac{\sigma_m^2\eta_\lambda^2 - \sigma_{\lambda m}^2\eta_\lambda\eta_m}{\sigma_\lambda^2\eta_m^2 + \sigma_m^2\eta_\lambda^2 - 2\sigma_{\lambda m}^2\eta_\lambda\eta_m}, \nn \\
\alpha_m &= \frac{\sigma_\lambda^2\eta_m^2 - \sigma_{\lambda m}^2\eta_\lambda\eta_m}{\sigma_\lambda^2\eta_m^2 + \sigma_m^2\eta_\lambda^2 - 2\sigma_{\lambda m}^2\eta_\lambda\eta_m}.
\end{align}
This is identical to the estimator constructed by \citet{Schmidt:2011ju}, but here for convergence rather than surface density. Note however that it is \emph{not} the maximum-likelihood estimator for a cut bivariate Gaussian as suggested by \citet{Schmidt:2011ju}. The likelihood \eqref{likelihood_hard_cuts} for a bivariate Gaussian size-magnitude distribution with hard cuts in size and magnitude has a normalization term in the denominator that depends on $\kappa$. The estimator derived above maximizes the exponent in the Gaussian (with some correction for the responsivities), but is not the maximum-likelihood estimator assuming a cut bivariate Gaussian size-magnitude distribution due to the omitted $\kappa$-dependent normalization. Nonetheless, provided one can calibrate $\eta_\lambda$ and $\eta_m$ from the data, taking a linear combination of the shift in mean size and magnitude as in Eqs. \eqref{linear_combination}-\eqref{weights} provides a simple unbiased point estimator for $\kappa$ using size and magnitude information.

To determine the responsivities $\eta_\lambda$ and $\eta_m$, in general one must have a model for the intrinsic distributions of sizes and magnitudes as well as detailed knowledge of the selection function. However, in the limit where we have hard cuts in size and magnitude (rather than a smooth selection function) and the sample is complete down to these limits, it is straightforward to estimate the responsivities directly from the data by estimating the derivative of the mean log-size and magnitude (above the cuts) with respect to $\kappa$, at a fiducial value of $\kappa = 0$.

\section{The joint size-magnitude distribution for CFHTLenS galaxies}
\label{joint_dist}
As we saw in the previous section, in order to write down the likelihood for $\kappa$ given the observed size, magnitude and redshift of a source we need to know the distribution of intrinsic (unlensed) sizes and magnitudes (as a function of redshift). How well we can recover $\kappa$ depends critically on the width of this distribution. In this section we build a simple model for the joint size-magnitude distribution, empirically motivated by the data, and fit this model to galaxies in CFHTLenS data in a number of narrow photometric redshift bins. 

\subsection{Parameterizing the size-magnitude distribution}
\label{joint_dist_param_model}
In order to fit the size-magnitude distribution, $p(m, \lambda)$, we must construct a parametrized model. We can conveniently divide the joint distribution into two parts by writing $p(m, \lambda) = p(\lambda | m)p(m)$. The Schechter function (or Gamma-distribution) has become ubiquitous in astronomy for describing the luminosity function of galaxies \citep{Schechter1976} and provides a remarkably robust description of the data. As such, we parametrize $p(m)$ as:
\begin{align}
\label{schechter}
p(m)dm \propto 10^{-0.4(\alpha + 1)(m-m_*)}\exp\left[-10^{-0.4(m-m_*)}\right]dm. 
\end{align}
As for the distribution of sizes conditional on magnitude, we choose a log-normal distribution (i.e., log-size $\lambda$ is normally distributed): 
\begin{align}
\label{log-normal}
p(\lambda|m) = \f{1}{\sqrt{2\pi}\sigma_\lambda(m)}\exp\left\{-\f{1}{2}\left[\f{\lambda - \langle\lambda\rangle(m)}{\sigma_\lambda(m)}\right]^2\right\}
\end{align}
where the mean-log-size $\langle\lambda\rangle(m)$ and dispersion $\sigma_\lambda(m)$ are both functions of magnitude. This is motivated both by previous successful applications in the literature \citep[e.g.,][]{Shen2003}, as well as indications from N-body simulations that suggest a log-normal size distribution may be expected \citep{Warren:1992hx, Cole:1995vh, Lemson:1999dl}. For the mean-log-size magnitude relation, we choose a parameterized form for $\langle\lambda\rangle(m)$ that corresponds to a power-law size-luminosity relation (i.e., $\langle\lambda\rangle(m)$ is linear in $m$). The dispersion is less well studied and in the absence of physical or empirical motivation we resort to choosing a parameterized function that apparently matches the data. We find that choosing the same functional form for both $\sigma_{\lambda}(m)$ and $\langle\lambda\rangle(m)$ provides a sufficiently close description of the data for our current purposes (see Fig. \ref{fig:size-mag-late}), although there is scope for building better, physically-motivated models for the joint size-magnitude distribution. The final parametrization of the size-magnitude distribution is given by Eqs. \eqref{schechter}-\eqref{log-normal} with:
\begin{align}
&\langle\lambda\rangle(m) = a_1 m + a_2,\nn \\
&\sigma_{\lambda}(m) = b_1 m + b_2.
\end{align}
This model has six free parameters: $\boldsymbol{\theta} = (\alpha,\,m_*,\,a_1, \,a_2,\, b_1,\, b_2)$. The likelihood for observing a galaxy with a particular size $\lambda$ and magnitude $m$ under this model is given by
\begin{align}
\label{size-mag_dist_model_full}
p(m, \lambda|\boldsymbol{\theta}) \propto &\f{1}{\sqrt{2\pi}\sigma_\lambda(m;\boldsymbol{\theta})}\exp\left\lbrace-\f{1}{2}\left[\f{\lambda - \langle\lambda\rangle(m;\boldsymbol{\theta})}{\sigma_\lambda(m;\boldsymbol{\theta})}\right]^2\right\rbrace \nn \\
&\times 10^{-0.4(\alpha + 1)(m-m_*)}\exp\left[{-10^{-0.4(m-m_*)}}\right].
\end{align}

\subsection{Significance of galaxy type}
We expect the intrinsic size-magnitude distribution to be different for different galaxy morphologies \citep[see e.g.,][]{Shen2003}. The distribution of galaxy sizes and magnitudes will hence be a mixture of different distributions, with each galaxy type contributing to this mixture in proportion to its relative abundance. We also expect that the relative abundance of different morphological types will be a function of local matter density --- the morphology-density relation \citep{Dressler1980, Postman1984, Whitmore1993, Hashimoto1999, Goto2003}. If we naively consider one global population of galaxies with all morphological types included, we would then expect the intrinsic distribution to vary considerably with local density, introducing a systematic effect not dissimilar to intrinsic alignments in the case of shear. This effect can be straightforwardly removed provided we have information on galaxy type available. The extent to which this effect can be eliminated will depend strongly on how reliably we can separate galaxies in a weak lensing survey by type. Here we attempt to remove the major component of this effect by splitting the galaxy population into two: a late-type and an early-type sample, defined principally by their photometrically-determined spectral type. We find that splitting the sample more finely by spectral type does not change the distributions significantly, but leave a more detailed exploration of this to future work.
\subsection{CFHTLenS sample}
\label{cfhtlens_sample}
The CFHTLenS survey covers 154 square degrees and is optimized for weak lensing measurements. The data and catalogue products are described in \citet{Erben2013, Heymans2012, Miller2013, Hildebrandt2012}. In this analysis we use all four wide fields (W1, W2, W3, W4) and take the catalogues presented in \citet{Erben2013}; namely galaxy shapes and sizes described in \citet{Miller2013} created using lens\emph{fit} \citep{Miller2007, Kitching2008}, and the photometric redshift measurements described in \citet{Hildebrandt2012}, created using the Bayesian photo-$z$ code BPZ \citep{Benitez:2000jr}.

Galaxy sizes (in our case image areas) are computed by combining the lens\emph{fit} scalelength $r_s$ (in units of arcseconds) and ellipticity $\epsilon$ for each object,
\begin{align}
\lambda = \ln\left(r_s\sqrt{\f{1-|\epsilon|}{1+|\epsilon|}}\,\right)
\end{align}
and we use $i$-band magnitudes throughout. We use the image masks described in \citet{Erben2013} (removing galaxies with $\mathrm{MASK} \geq 1$, as described in that paper), only galaxies with a lens\emph{fit} weight larger than zero are included, and the redshift range is restricted to $0.4 < z_p < 1.3$, selected by peak posterior redshift $z_p$ provided by BPZ. The sample is divided into two by spectral type, based on the maximum likelihood spectral type $T_\mathrm{B}$ provided by BPZ and the lens\emph{fit} bulge-fraction $b$, with the late-type sample defined by $T_B > 2$ and $b < 1$ and the early-type sample with $T_\mathrm{B} \leq 2$ and $b > 0.25$.  The late-type sample contains $\num[group-separator={,}]{2771149}$ galaxies, whilst the early-type sample contains $\num[group-separator={,}]{151724}$. The sub-samples are further divided into photometric redshift bins selected by their peak posterior photo-$z$. The late-type sample is divided into 9 equally-spaced photometric redshift bins between $z_p=0.4$ and $z_p=1.3$. The smaller early-type is divided into 3 bins of $0.4 < z_p < 0.6$, $0.6 < z_p < 0.8$ and $0.8 < z_p < 1.3$ to ensure enough galaxies per bin to allow for robust estimation of the intrinsic size-magnitude distribution.

Note that in the galaxy-type split we remove galaxies with apparently contradictory morphology indicators from BPZ and lens\emph{fit}. We discard sources indicated to be late-type by their BPZ maximum-likelihood spectral type but given a bulge-fraction equal to one by lens\emph{fit} (in practise this removes a very small fraction of sources). We also also cut galaxies indicated to be early-type by BPZ but given a small bulge-fraction ($<0.25$) by lens\emph{fit}. Including all sources with $T_\mathrm{B} \leq 2$ gives an empirical size-magnitude distribution that appears to be a mixture between distinct (overlapping) populations; this is indicative of either sub-populations with different size-magnitude distributions, or contamination from mis-classified late-types. The additional cut on bulge-fraction appears to mitigate this issue, but for future work optimal galaxy-type separation should be investigated. Less than $10\%$ of the sample is discarded due to the galaxy-type split adopted here.

\subsection{Simplifying assumptions}
\begin{figure}
\includegraphics[width = 8.22cm]{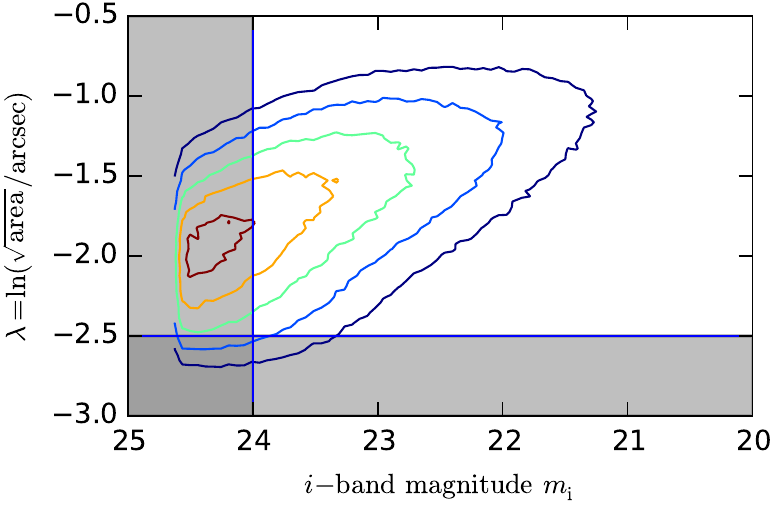}
\caption{Iso-probability contours for the 2D histogram of the CHFTLenS galaxy sample prior to the hard magnitude and size cuts at $m_i<24$ and $\lambda > -2.5$ respectively, above which the sample is assumed to be complete and found to be well described by a Schechter luminosity function and log-normal size distribution (at fixed magnitude). Contours are drawn at 90, 70, 50, 30 and 15\% of the peak probability-density.}
\label{fig:selection}
\end{figure}

For the purposes of estimating the intrinsic size-magnitude distribution we make a number of simplifying assumptions. We make hard cuts in size and magnitude in the regime where the data can be assumed to be complete, to avoid modelling and measuring the selection function simultaneously with the size-magnitude distribution. We make a hard magnitude cut at $m_i<24$, and a size-cut at $\lambda = \ln(\sqrt{\mathrm{Area}}/\mathrm{arcsec}) > -2.5$ above which we assume the sample to be complete (see Fig. \ref{fig:selection}) and find that the size-magnitude distribution is well described by a Schechter luminosity function and log-normal size distribution conditional on magnitude in this region. These cuts remove roughly $1/3$ of the available sources. This highlights the importance of modelling the selection function allowing us to use the full sample. For our current purposes we fit the joint distribution in the complete regime and later (in \S \ref{smooth_selection_fits}) impose an approximate smooth selection function to study its impact. We assume that the lensing effect averages to zero over the whole sky, so the joint size-magnitude distribution measured over a sufficiently large patch of sky and large number of sources will be close to the intrinsic size-magnitude distribution. In practice, the observed distribution of sizes and magnitudes will be the intrinsic distribution convolved with the convergence distribution $p(\kappa)$. Provided that $\langle\kappa\rangle=0$, this convolution with $p(\kappa)$ will preserve the mean of the distribution but will broaden it (and also impact higher moments). However, the effect will be small since the width of $p(\kappa)$ will be much smaller than either $\sigma_\lambda$ or $\sigma_m$, and can in any case be corrected for. We ignore this effect here.
\subsection{Parameter estimation}
We fit the parametrized model described in \S \ref{joint_dist_param_model} to the data in each photo-$z$ bin (and for each type) using MCMC Metropolis-Hastings with wide uniform priors on all parameters. Multiple chains were run in each case and a Gelman-Rubin test \citep{Gelman1992} was performed to indicate convergence (ensuring that the Gelman-Rubin statistic $R < 1.03$ in every case).

\begin{figure*}
\includegraphics[width = 17.7cm]{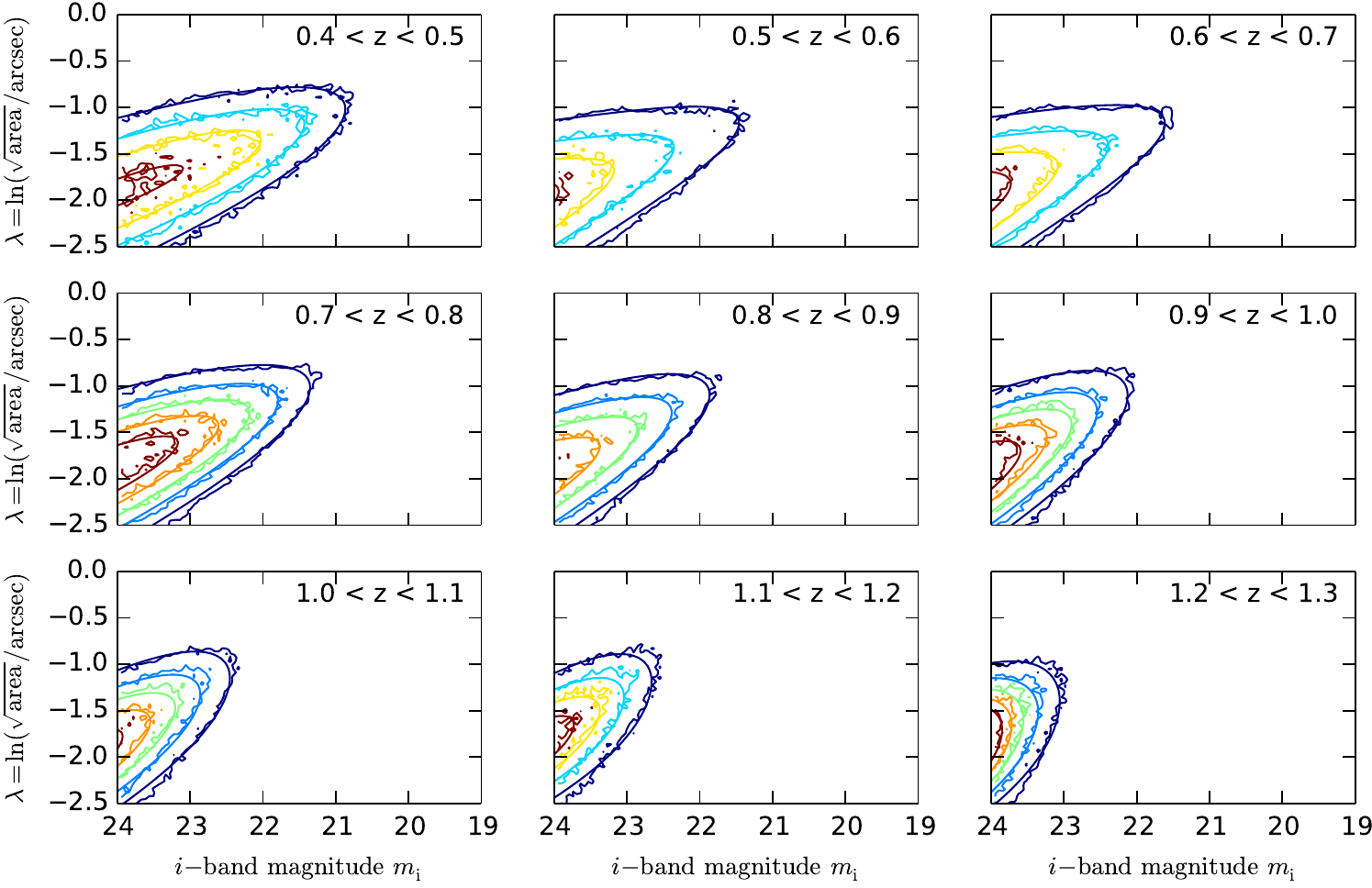}
\caption{Size-magnitude distribution for a number of narrow photometric redshift bins for the late-type sample. The wiggly contours are iso-probability-density contours for the 2D histogram of the data, whilst the smooth curves are the model fits to the data (taking the expected marginalized posterior parameter values). Contours are drawn at 80, 60, 40 and 20\% of the peak probability-density.}
\label{fig:size-mag-late}
\end{figure*}

\begin{figure*}
\includegraphics[width = 17.7cm]{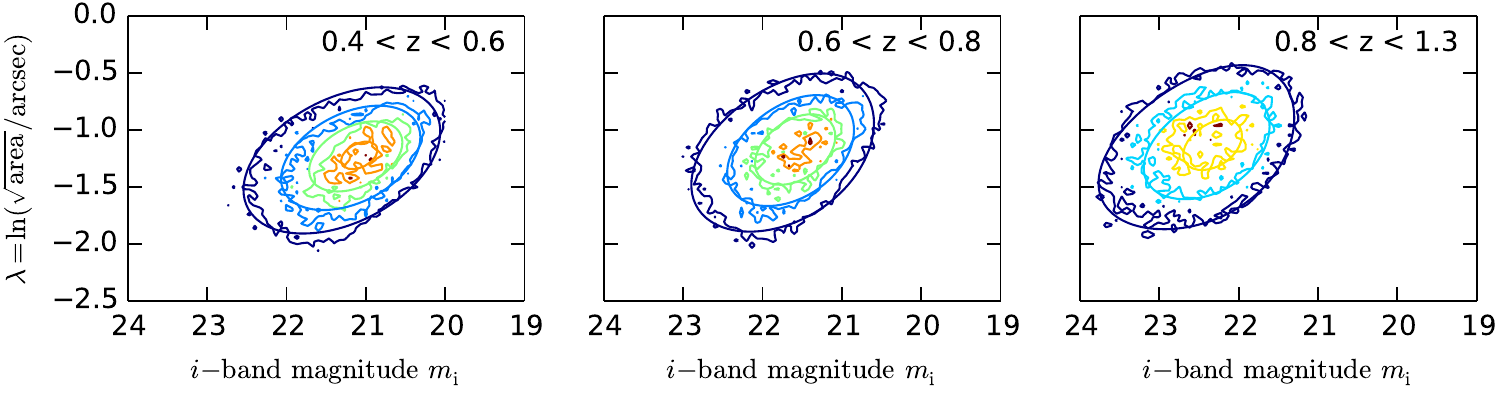}
\caption{Size-magnitude distribution for a number of narrow photometric redshift bins for the early-type sample. The wiggly contours are iso-probability-density contours for the 2D histogram of the data, whilst the smooth curves are the model fits to the data (taking the expected marginalized posterior parameter values). Contours are drawn at 80, 60, 40 and 25\% of the peak probability-density.}
\label{fig:size-mag-early}
\end{figure*}

Figs. \ref{fig:size-mag-late} and \ref{fig:size-mag-early} show the 2D histograms of size-magnitude data and the respective fitted distributions (taking the expected marginal parameter values) for CFHTLenS galaxies for the late- and early-type samples respectively. Table \ref{tab:joint_dist_params} gives the expected marginal parameter values and their standard deviations. The model gives a good description of the data, but there is clearly room for improvement and we highlight the development of more sophisticated models of the size-magnitude distribution as important future work. There is evidently some evolution of the model parameters with redshift. In a more sophisticated set-up one could build an extended model for the 3D size-magnitude-redshift distribution and fit the entire data-set over the full range of redshift.  


For comparison, we also compute the calibration parameters $\alpha_\lambda,\;\alpha_m\;\eta_\lambda$ and $\eta_m$ for the estimator developed in \S \ref{estimator} for each subset of the data as described above. The responsivities are computed by estimating the derivative of $\langle\lambda\rangle$ and $\langle m \rangle$ with respect to $\kappa$ (at a fiducial value of $\kappa = 0$). We provide bootstrap errors on all calibration parameters computed from 1000 bootstrap samples. These results are summarised in Table \ref{tab:calibration_params}.
\begin{table*}
\caption{Expected (marginalized) posterior parameter values with associated $1\sigma$ uncertainties (in the last significant figure) for the late- and early-type samples.}
\begin{tabular}{ccccccc}
\hline
Photo-z bin & $\alpha$ & $m_*$ & $a_1$ & $a_2$ & $b_1$ & $b_2$\\
\hline \\
& & &  Late-type & & &  \\ \\
$0.4 < z < 0.5$ & $-$1.019(6) & 21.58(1) & $-$0.287(1) & 4.98(2) & 0.0542(9) & $-$0.83(2)\\
$0.5 < z < 0.6$ & $-$1.429(6) & 21.33(1) & $-$0.299(1) & 5.24(2) & 0.0601(7) & $-$0.97(2)\\
$0.6 < z < 0.7$ & $-$1.436(5) & 21.63(1) & $-$0.3086(9) & 5.5(2) & 0.0597(7) & $-$0.97(2)\\
$0.7 < z < 0.8$ & $-$1.042(7) & 22.248(8) & $-$0.328(1) & 5.98(2) & 0.0597(8) & $-$0.98(2)\\
$0.8 < z < 0.9$ & $-$1.174(8) & 22.511(9) & $-$0.368(1) & 6.96(3) & 0.0513(9) & $-$0.8(2)\\
$0.9 < z < 1.0$ & $-$1.12(1) & 22.96(1) & $-$0.406(2) & 7.88(4) & 0.04(1) & $-$0.54(3)\\
$1.0 < z < 1.1$ & $-$0.97(2) & 23.37(1) & $-$0.427(2) & 8.44(5) & 0.019(2) & $-$0.06(3)\\
$1.1 < z < 1.2$ & $-$0.63(3) & 23.76(2) & $-$0.433(3) & 8.63(8) & $-$0.009(2) & 0.61(6)\\
$1.2 < z < 1.3$ & $-$1.73(5) & 23.8(3) & $-$0.387(5) & 7.5(1) & $-$0.026(4) & 1.04(9)\\
\hline \\
& & &  Early-type & & & \\ \\
$0.4 < z < 0.6$ & 1.24(1) & 22.021(8) & $-$0.208(2) & 3.15(5) & 0.05(2) & $-$0.7(3)\\
$0.6 < z < 0.8$ & 1.51(2) & 22.594(8) & $-$0.236(3) & 3.92(6) & 0.039(2) & $-$0.45(4)\\
$0.8 < z < 1.3$ & 0.96(2) & 23.07(1) & $-$0.233(3) & 4.07(7) & 0.022(2) & $-$0.06(5)\\
\hline
\end{tabular}
\label{tab:joint_dist_params}
\end{table*}
\begin{table*}
\caption{Calibration parameters for the estimator derived in \S \ref{estimator} for the late and early type samples with bootstrap ($1 \sigma$) errors computed from 1000 bootstrap samples.}
\begin{tabular}{cccccccc}
\hline
Photo-z bin & $\sigma_\lambda$ & $\sigma_m$ & $\sigma_{\lambda m}^2$ & $\eta_\lambda$ & $\eta_m$ & $\alpha_\lambda$ & $\alpha_m$\\
\hline \\
& & & Late-type & & & & \\ \\
$0.4 < z < 0.5$ & 0.44(1) & 0.88(1) & $-$0.2(1) & 0.57(2) & $-$0.8(4) & 0.82(3) & 0.18(3)\\
$0.5 < z < 0.6$ & 0.436(3) & 0.84(2) & $-$0.185(8) & 0.45(6) & $-$0.5(1) & 0.93(6) & 0.07(6)\\
$0.6 < z < 0.7$ & 0.433(4) & 0.78(4) & $-$0.16(1) & 0.47(5) & $-$0.6(1) & 0.81(6) & 0.18(6)\\
$0.7 < z < 0.8$ & 0.428(5) & 0.72(6) & $-$0.15(2) & 0.52(5) & $-$0.8(1) & 0.5(1) & 0.5(1)\\
$0.8 < z < 0.9$ & 0.43(5) & 0.65(8) & $-$0.14(2) & 0.46(5) & $-$0.7(1) & 0.5(2) & 0.5(2)\\
$0.9 < z < 1.0$ & 0.431(5) & 0.6(1) & $-$0.12(3) & 0.36(7) & $-$0.7(1) & 0.2(3) & 0.8(3)\\
$1.0 < z < 1.1$ & 0.432(4) & 0.5(1) & $-$0.1(3) & 0.33(8) & $-$0.7(1) & 0.1(3) & 0.9(3)\\
$1.1 < z < 1.2$ & 0.422(5) & 0.5(1) & $-$0.09(4) & 0.34(8) & $-$0.8(1) & 0.0(3) & 1.0(3)\\
$1.2 < z < 1.3$ & 0.417(7) & 0.4(2) & $-$0.05(5) & 0.3(1) & $-$0.5(1) & 0.1(3) & 0.9(3)\\
\hline \\
& & & Early-type & & & &  \\ \\
$0.4 < z < 0.6$ & 0.382(1) & 0.781(3) & $-$0.112(1) & 0.96(1) & $-$2.1(2) & 0.45(1) & 0.55(1)\\
$0.6 < z < 0.8$ & 0.42(2) & 0.71(4) & $-$0.111(1) & 0.94(2) & $-$2.08(2) & 0.3(8) & 0.7(8)\\
$0.8 < z < 1.3$ & 0.44(3) & 0.68(4) & $-$0.104(4) & 0.92(2) & $-$1.97(6) & 0.26(8) & 0.74(8)\\
\hline
\end{tabular}
\label{tab:calibration_params}
\end{table*}
\subsection{Selection function}
\label{smooth_selection_fits}
\begin{figure*}
\includegraphics[width = 17.7cm]{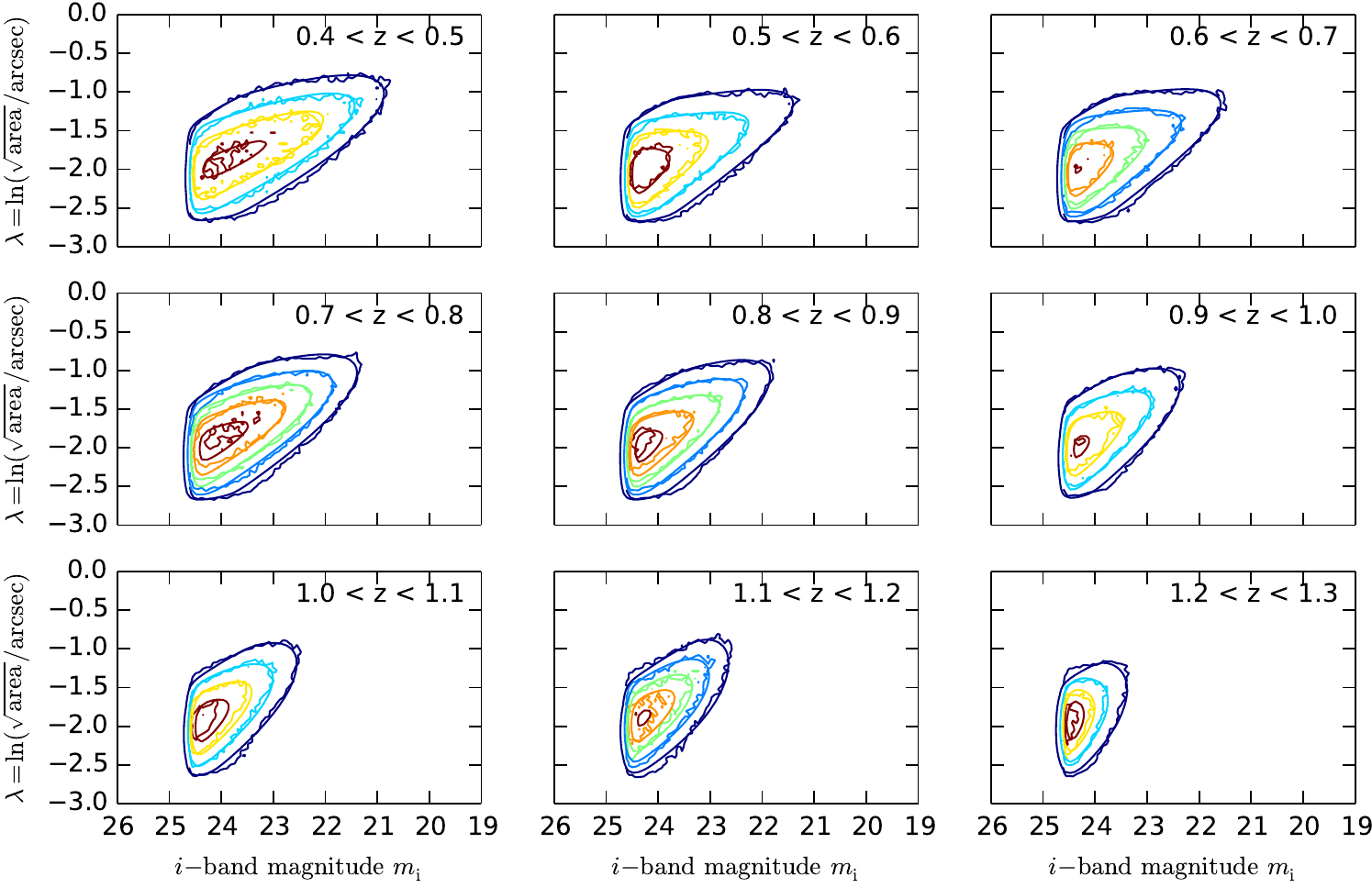}
\caption{Size-magnitude distribution for a number of narrow photometric redshift bins for the late-type sample, without size and magnitude cuts. The wiggly contours are iso-probability-density contours for the 2D histogram of the data, whilst the smooth curves are the model fits to the data (taking the expected marginalized posterior parameter values) multiplied by the selection function described in Eq. \eqref{select}. Contours are drawn at 80, 60, 40 and 20\% of the peak probability-density.}
\label{fig:size-mag-smooth}
\end{figure*}

As mentioned previously, the hard size and magnitude cuts imposed so far to ensure a complete sub-sample remove $\sim1/3$ of the available sources. In practice we want to use all sources with size and magnitude measurements, and must consequently have a model for the selection function. For illustrative purposes we impose an approximate smooth selection function that, combined with the fitted model for the intrinsic size-magnitude distribution, gives a rough description of the data well into the incomplete regime. The selection function can be approximated by
\begin{align}
\label{select}
S(m, \lambda) = \frac{1}{8}&\left[1 + \mathrm{tanh}\left(-\f{m - m_c}{s_m}\right)\right]\left[1 + \mathrm{tanh}\left(\frac{\lambda - \lambda_c}{s_\lambda}\right)\right] \nn \\
&\times\left[1 + \mathrm{tanh}\left(\frac{m/2.17 + \lambda - \mu_c}{s_\mu}\right)\right],
\end{align}
where the first bracket represents selection due to a magnitude limit, the second due to a size limit, and the third due to a surface brightness limit. We take $m_c = 24.6$, $s_m = 0.15$, $\lambda_c = -2.67$, $s_\lambda = 0.07$, $\mu_c = 9.9$ and $s_\mu = 0.05$, chosen to give a rough description for the data sufficient for our current purposes (no formal fit was made). Fig. \ref{fig:size-mag-smooth} shows the 2D histograms of the data in each redshift bin and the model fits to the size-magnitude distribution multiplied by the selection function in Eq. \eqref{select} (for the late-type sub-samples only, since the impact of the selection function on the early-types is much smaller). In a more careful analysis one should fit the parameterized selection function simultaneously with the intrinsic size-magnitude distribution.
\section{How well can we estimate $\kappa$ from sizes and magnitudes?}
\label{dispersion}
Given the intrinsic joint distribution of sizes and magnitudes we would like to know how well we expect to be able to estimate $\kappa$. In \S \ref{magnification_estimation}, we provided two approaches to estimating $\kappa$ from a given size and magnitude: a Bayesian approach where we derived the posterior distribution for $\kappa$ given a measured size and magnitude, as well as a simple unbiased (but sub-optimal) point estimator based on the shift in average size and magnitude. We would like to estimate how well we could recover $\kappa$ from the data from both the likelihoods derived in \S \ref{bayesian} and the estimator described in \S \ref{estimator}. For the latter, we simply compute the variance of the estimator in each redshift bin; these results are summarized in Table \ref{tab:fisher_errors}.

For the Bayesian approach, we leave the development of a globally Bayesian framework for cosmological parameter inference incorporating the full posterior information on $\kappa$ for every source to later work. Rather, we estimate an effective dispersion on $\kappa$ estimation from the likelihoods derived in \S \ref{bayesian}; we compute the Fisher Information for $\kappa$ from the likelihood and obtain an estimate of the dispersion on $\kappa$ via the Cram\'{e}r-Rao inequality. The Fisher Information matrix $F_{\alpha \beta}$ is the expectation of the second derivative of the negative log-likelihood with respect to the model parameters $\alpha$ and $\beta$:
\begin{align}
F_{\alpha \beta} = -\left\langle\f{\partial^2\ln L}{\partial\theta_\alpha\partial\theta_\beta}\right\rangle.
\end{align}
The Cram\'{e}r-Rao inequality allows us to obtain a lower bound on the marginal errors of the model parameters via the diagonal elements of the inverse Fisher matrix:
\begin{align}
\sigma_\alpha \geq\sqrt{(F^{-1})_{\alpha\alpha}}.
\end{align}
Here we have a single parameter, $\kappa$, so the Fisher Information gives us an estimate of the anticipated uncertainty in $\kappa$ estimation from sizes and magnitudes:
\begin{align}
\sigma_\kappa = \sqrt{1/F_{\kappa\kappa}}.
\end{align}
In the limit where a large number of sources is used to make an inference about a single value of $\kappa$, the likelihood is asymptotically Gaussian with variance $1/F_{\kappa\kappa}$. This variance will dominate the noise in $\kappa$ estimation. Note that the Fisher Information only involves the likelihood and is not an explicitly Bayesian quantity, but nonetheless allows us to estimate the uncertainty on convergence estimation from the likelihood of interest.

We compute the Fisher Information for the likelihoods derived in Eq. \eqref{likelihood_hard_cuts} and \eqref{likelihood_smooth_selection}, taking the parameterized fits to the joint size-magnitude distribution described in Eq. \eqref{size-mag_dist_model_full} and Table \ref{tab:joint_dist_params}. We do not marginalize over uncertainties in the size-magnitude distribution parameters; the model parameters are sufficiently tightly constrained that this effect should be small, but a more careful analysis should formally marginalize over all model parameters.

Table \ref{tab:fisher_errors} shows both the Fisher Information forecast uncertainties in $\kappa$ estimation and the variance of the point estimator from \S \ref{estimator}, from the fitted joint size-magnitude distribution in the redshift bins described in \S \ref{cfhtlens_sample}. The estimated dispersion on convergence from the likelihoods derived in \S \ref{bayesian} and the estimator from Eqs. \eqref{linear_combination}-\eqref{weights} are largely consistent (within the uncertainties in the latter), with a tendency for the likelihoods from \S \ref{bayesian} that account for the full shape of the size-magnitude distribution to outperform the sub-optimal estimator, as one might expect.
\begin{table*}
\caption{Fisher forecast errors on $\kappa$ from the likelihood of equation \eqref{likelihood_hard_cuts} for the joint size-magnitude distribution fits summarized in Figs. \ref{fig:size-mag-late} - \ref{fig:size-mag-early} and Table \ref{tab:joint_dist_params}, and the variance of the calibrated estimator developed in \S \ref{estimator} for the same sub-samples.}
\begin{tabular}{cccc}
\toprule
Photo-z bin & $\sigma_\kappa = \sqrt{1/F_{\kappa\kappa}}$ (hard cuts) & $\sigma_\kappa = \sqrt{1/F_{\kappa\kappa}}$ (smooth selection function) & $\sigma_{\hat{\kappa}}$ (estimator)\\
\midrule \\& &Late-type &   \\ \\
$0.4 < z < 0.5$ & 0.68 & 0.72 & 0.76(2)\\
$0.5 < z < 0.6$ & 0.81 & 0.85 & 1.0(1)\\
$0.6 < z < 0.7$ & 0.82 & 0.89 & 0.89(9)\\
$0.7 < z < 0.8$ & 0.70 & 0.74 & 0.7(1)\\
$0.8 < z < 0.9$ & 0.79 & 0.85 & 0.81(9)\\
$0.9 < z < 1.0$ & 0.78 & 0.85 & 0.81(9)\\
$1.0 < z < 1.1$ & 0.73 & 0.78 & 0.72(9)\\
$1.1 < z < 1.2$ & 0.63 & 0.65 & 0.6(1)\\
$1.2 < z < 1.3$ & 0.87 & 0.99 & 0.8(1)\\

\midrule \\& &Early-type &   \\ \\
$0.4 < z < 0.6$ & 0.3 & -- & 0.31(3)\\
$0.6 < z < 0.8$ & 0.3 & -- & 0.31(4)\\
$0.8 < z < 1.3$ & 0.35 & -- & 0.32(5)\\
\bottomrule
\end{tabular}
\label{tab:fisher_errors}
\end{table*}
The typical effective dispersion on $\kappa$ estimation is $\sigma_\kappa = 0.8$; this is the weighted mean of $\sigma_\kappa$ for late-types in each redshift bin, including the smooth selection function, and the early-types in their respective redshift bins. Comparing to shear, for which the dispersion in (complex) ellipticities is around $\sigma_e = 0.38$, we see that magnification will clearly be statistically less powerful than cosmic shear (notwithstanding systematics). Comparing the Fisher forecast errors for the case with hard cuts to the case of a smooth selection function, we find similar values for the dispersion with the latter generally giving a slightly higher dispersion in $\kappa$ (although this discrepancy is small compared to the increase in numbers achieved by including the full source sample). The additional sources added by extending into the incomplete region have moderately larger intrinsic scatter in their sizes, since $\sigma_\lambda(m)$ is a (slowly) increasing function of magnitude. This accounts for at least some of the difference between the two scenarios. Note that whilst the extended distribution (with smooth selection) is `broader' than its brother with hard cuts, that does not necessarily imply that the likelihood derived from this distribution will be less strongly peaked; there is a competition between the width of the distribution and the effect of pushing sources under the selection function both impacting the posterior distribution for $\kappa$.

Some of the error budget on convergence estimation will come from photo-$z$ errors and uncertainties in size and magnitude measurement. We believe both of these effects to be small. Photometric redshift errors will contribute directly to the scatter in sizes and magnitudes. The observed size-magnitude distributions for sources binned in photo-$z$ can be thought of as the redshift-evolving size-magnitude distribution, smoothed over some kernel in redshift (for example a Gaussian kernel for Gaussian photo-$z$ errors). We obtain a crude estimate the effect of photo-$z$ errors on $\sigma_\kappa$ as follows. We constructed a redshift-evolving size-magnitude distribution assuming linear redshift evolution in all of the model parameters (given in Table \ref{tab:joint_dist_params} for each redshift bin). We find that smoothing this redshift-evolving size-magnitude distribution with a Gaussian kernel with dispersion $\sigma_z = 0.05(1+z)$ has a small effect; for example, typically the dispersion in log-size as a function of magnitude is broadened by $\lesssim 3\%$. The measurement uncertainties on magnitude and size are both small compared to $\sigma_m$ and $\sigma_\lambda$ so should not have a significant impact on $\sigma_\kappa$.

It may be possible to reduce the dispersion on convergence estimation by exploiting tight scaling relations between quantities that change under magnification (e.g., size and magnitude) and quantities that are conserved under lensing. For example, \citet{Bertin2006} and \citet{Huff2014} describe using the well-known fundamental plane \citep{Dressler1987, Djorgovski1987} to obtain tight convergence estimates for early-type galaxies. Similar scaling relations may be useful for late-types; this should be explored in future work.

In \S \ref{fisher} we explore the implications of our results for the dispersion on convergence estimation in terms of cosmological parameter inference, in different scenarios with respect to systematics.

\section{Forecasts for joint magnification-cosmic shear analysis}
\label{fisher}
In this section we explore what improvements might be expected for cosmological parameter estimation by combining cosmic shear with magnification. We extend the work of HAJ13 to include: (1) updated information on the effective dispersion of a plausible convergence estimator from sizes and magnitudes, and (2) forecasts including systematic effects.

Suppose we have a point estimator for shear $\hat{\gamma} = \hat{\gamma}_1 + i\hat{\gamma}_2 = \epsilon$ (i.e., the complex ellipticity of each source) and a similar estimator for convergence $\hat{\kappa}$. The latter could be, for example, the maximum likelihood estimator from the likelihood developed in \S \ref{bayesian}, the estimator developed in \S \ref{estimator} or indeed some other estimator. Following HAJ13, in the absence of intrinsic alignments or intrinsic size-magnitude correlations the tomographic power spectra between tomographic bins $i$ and $j$ of these estimators are given by:
\begin{align}
\hat{C}^{\kappa\kappa}_{\ell,ij} &= C^\mathrm{GG}_{\ell,ij} + \delta_{ij}\sigma_{\kappa}^2/\bar{n}_i \nn \\
\hat{C}^{\gamma_1\gamma_1}_{\ell,ij} &= \hat{C}^{\gamma_2\gamma_2}_{\ell,ij} = C^\mathrm{GG}_{\ell,ij}+ \delta_{ij}\sigma_{\gamma}^2/\bar{n}_i \nn \\
\hat{C}^{\gamma_2\gamma_1}_{\ell,ij} &= C^\mathrm{GG}_{\ell,ij} \nn \\
\hat{C}^{\gamma_1\kappa}_{\ell,ij} &= \hat{C}^{\gamma_2\kappa}_{\ell,ij} = C^\mathrm{GG}_{\ell,ij},
\end{align}
where $\sigma_\kappa$ and $\sigma_\gamma$ are the intrinsic dispersions in the convergence and shear estimators, $\bar{n}_i$ is the number density of galaxies in the $i$-th bin and $C^\mathrm{GG}_{\ell,ij}$ are the tomographic lensing power spectra, which in the \citet{Limber1954} approximation are given by \citep{Takada2004}:
\begin{align}
C^\mathrm{GG}_{\ell,ij} &= \int\f{d\chi}{\chi_\mathrm{m}^{2}(\chi)}w_{i}(\chi)w_{j}(\chi)\left[1+z(\chi)\right]^2P(k;\chi),
\end{align}
where $\chi$ is comoving distance, $P(k; \chi)$ is the 3D matter power spectrum, $k = \ell/\chi_\mathrm{m}(\chi)$ and $\chi_\mathrm{m}(\chi)$ is the transverse comoving distance corresponding to comoving distance 
$\chi$. The lensing weight functions $w_{i}(\chi)$ are given by
\begin{align}
w_{i}(\chi)=\f{3\Omega_\mathrm{m}H_0^2}{2}\chi_\mathrm{m}(\chi)\int_\chi^{\chi_\mathrm{H}} d\chi'\;n_i(\chi')\f{\chi_\mathrm{m}(\chi'-\chi)}{\chi_\mathrm{m}(\chi')}, 
\end{align}
where $n_i(\chi)d\chi$ is the galaxy redshift distribuiton for the $i$-th tomographic bin. 
\subsection{Intrinsic alignments (IAs)}
\label{sec:intrinsic_alignments}
Nearby galaxies form in a similar tidal gravitational field, which may lead to a preferred ellipticity orientation of neighbouring galaxies. We can model this intrinsic alignment (IA) as an additional contribution to the observed shape of a galaxy, i.e. $\epsilon = \gamma^G + \gamma^I + \epsilon_s$, where $\gamma^G$ is the gravitational lensing shear, $\gamma^I$ is the contribution to the intrinsic ellipticity due to the local environment and $\epsilon_s$ is the random (unlensed) shape of the galaxy drawn from the global distribution of intrinsic ellipticities (averaged over shear and intrinsic alignments). Intrinsic alignments introduce two additional terms to the ellipticity power spectra: the first is the intrinsic-intrinsic (II) correlation, which arises due to the correlation of ellipticities of nearby galaxies due to their shared local environment (and associated preferential orientation). The second effect arises from the cross-correlation between intrinsic ellipticity and cosmic shear (GI); background galaxies are lensed by foreground gravitational potentials which in turn govern the orientation of the foreground galaxies, leading to an anti-correlation between foreground and background (lensed) galaxies. The II correlation adds positively to the total measured shear signal while the GI term subtracts from it.

The II correlation has been detected for low redshift LRGs \citep{Brown2002, Heymans2004, Okumura2009b} and a consistent but null detection was made by \citet{Mandelbaum2006}. The GI correlation is somewhat easier to measure since the random ellipticity only appears once in the calculation and has been detected for LRGs at low redshift ($z < 0.5$)\citep{Mandelbaum2006, Hirata2007, Okumura2009} and at intermediate redshift $z\sim 0.5$ \citep{Joachimi2011}. For blue galaxies, \citet{Mandelbaum2011} do not detect significant II or GI at intermediate redshift $z\sim0.6$.

There are a number of approaches to mitigating the effects of IAs. The II contribution can be largely eliminated by removing or down-weighting nearby galaxies in the analysis \citep{Heymans2003, King2002, King2003, TakadaWhite2004}. The GI contribution is not localized in redshift and is as such more difficult to remove. A more sophisticated approach has been developed \citep{King:2005dg, Joachimi2008, Joachimi:2009id} whereby the GI and II signal can be ``nulled": by exploiting the different characteristic redshift dependence of the GG, II and GI power spectra it is possible to project the signal into a number of template functions and isolate the GG, II and GI contributions, at the cost of some signal-to-noise.

Alternatively, a model can be assumed for all IA contributions and they can be included in the analysis. Since the IA terms inevitably contain cosmological dependence, the choice of IA model will impact the cosmological parameter inference and poor modelling of IAs could lead to biased cosmological parameter estimation \citep{Kirk2012}. Modelling and simulating IAs is a formidable theoretical challenge and we may not be fortunate enough to have an IA model that we have sufficient confidence in to rely on modelling alone. In this situation we can construct a flexible IA model containing a number of free nuisance parameters which can be simultaneously estimated along with the cosmological parameters, and then marginalized over, in the hope of eliminating the sensitivity to the fiducial IA model at the cost of some signal-to-noise. This approach has been applied at the Fisher matrix level \citep{Bridle:2007il, Bernstein:2009jw, Joachimi:2009id, Kirk2012} and applied to data by \citet{Kirk2010}. We adopt this approach here.

The II and GI intrinsic alignment tomographic power spectra take the form:
\begin{align}
C^{\gamma^\mathrm{I}\gamma^\mathrm{I}}_{\ell,ij} = \int \f{d\chi}{\chi_\mathrm{m}(\chi)^2}\;n_{i}(\chi)n_{j}(\chi)P_{\gamma^\mathrm{I}\gamma^\mathrm{I}}(k; \chi) \nn \\
C^{G\gamma^\mathrm{I}}_{\ell,ij} = \int \f{d\chi}{\chi_\mathrm{m}(\chi)^2}\;w_{i}(\chi)n_{j}(\chi)P_{\delta\gamma^\mathrm{I}}(k; \chi)
\end{align} 
where $P_{\gamma^\mathrm{I}\gamma^\mathrm{I}}$ is the power spectrum of the intrinsic alignment contribution to the ellipticities, $P_{\delta\gamma^\mathrm{I}}$ is the cross-power between the intrinsic ellipticity and the matter density field and again it is understood that $k = \ell/\chi_\mathrm{m}(\chi)$. These are the central predictions of any intrinsic alignment model. We take as our fiducial IA model the linear alignment (LA) model of \citet{Hirata2004}, with II and GI power spectra:
\begin{align}
&P_{\gamma^\mathrm{I}\gamma^\mathrm{I}}(k;\chi) = (-C_1\rho_0)^2P^\mathrm{lin}_{\delta\delta}(k; \chi = 0), \nn \\
&P_{\delta\gamma^\mathrm{I}}(k;\chi) = -C_1\rho_0\sqrt{P^\mathrm{lin}_{\delta\delta}(k; \chi = 0)P_{\delta\delta}(k; \chi)},
\end{align} 
where $\rho_0$ is the matter density today, $C_1$ is the amplitude of the IA signal and $D(z)$ is the growth factor. We take as our fiducial value for the IA normalization $C_1 = 5 \times 10^{−14} (h^2 M_\odot \mathrm{Mpc}^{-3} )^{-1}$ following \citet{Bridle:2007il}. 

To parameterize our ignorance about the IA model, we introduce flexibility via a grid of nuisance parameters:
\begin{align}
\label{nuisance_grid}
b(k,z) = A\;Q(k,z)
\end{align}
where $A$ is some amplitude and $Q(k,z)$ is a spline interpolation function over a grid of $n_k\times n_z$ nodes (logarithmically spaced in $k$ and $z$), each of which is allowed to vary independently about a fiducial value of $1$. For more details of this nuisance parameter set-up see \citet{Kirk2012} and \citet{Joachimi:2009id}. The flexible intrinsic alignment II and GI power spectra then read
\begin{align}
&C^{\gamma^\mathrm{I}\gamma^\mathrm{I}}_{\ell,ij} = \int \f{d\chi}{\chi_\mathrm{m}(\chi)^2}\;n_{i}(\chi)n_{j}(\chi)b_{\gamma^\mathrm{I}}^2(k,z)P^\mathrm{lin}_{\delta\delta}(k; 0), \nn \\
&C^{G\gamma^\mathrm{I}}_{\ell,ij} = \int \f{d\chi}{\chi_\mathrm{m}(\chi)^2}\;w_{i}(\chi)n_{j}(\chi)r_{\gamma^\mathrm{I}}(k,z)b_{\gamma^\mathrm{I}}(k,z) \nn \\
&\;\;\;\;\;\;\;\;\;\;\;\;\;\;\;\;\;\;\;\;\;\;\;\;\;\;\times\sqrt{P^\mathrm{lin}_{\delta\delta}(k; \chi = 0)}\sqrt{P_{\delta\delta}(k; \chi)},
\end{align}
where $b_{\gamma^\mathrm{I}}(k,z)$ and $r_{\gamma^\mathrm{I}}(k,z)$ take the form of \eqref{nuisance_grid}.
\subsection{Intrinsic size-magnitude-density correlations (ISCs)}
\label{sec:isc}
In the case of size and magnitude information, we may expect the average size and/or magnitude of galaxies to depend on local density. These correlations of sizes and/or magnitudes with local density will give rise to an equivalent II and GI term in the estimated convergence power spectrum, where the II arises from the correlation of sizes and magnitudes of nearby sources due to their shared local environment, and the GI arises from the correlation of the sizes/magnitudes of foreground (lensing) galaxies with background (lensed) galaxies in a similar fashion to the GI term in the intrinsic alignment case. As with IAs, the intrinsic size/magnitude density correlations (ISCs) can be modelled as an additional contribution to the convergence estimator, i.e., $\hat{\kappa} = \kappa^\mathrm{G} + \kappa^\mathrm{I} + \kappa_\mathrm{s}$, where $\kappa^\mathrm{G}$ is the gravitation lensing convergence, $\kappa^\mathrm{I}$ is the systematic contribution due to ISCs and $\kappa_\mathrm{s}$ is the random component drawn from the global distribution of $\hat{\kappa}$. For a joint magnification-cosmic shear analysis, we must also consider the cross correlation of the shear and convergence estimators. With both intrinsic alignments and intrinsic size/magnitude correlations, there will be a cross-correlation between the intrinsic alignment contribution to the shear and the size/magnitude-density contribution to the convergence estimator. The tomographic power spectra for the additional intrinsic terms due to size/magnitude density correlations will take the form (similar to IAs)
\begin{align}
C^{\kappa^\mathrm{I}\kappa^\mathrm{I}}_{\ell,ij} = \int \f{d\chi}{\chi_\mathrm{m}(\chi)^2}\;n_{i}(\chi)n_{j}(\chi)P_{\kappa^\mathrm{I}\kappa^\mathrm{I}}(k; \chi), \nn \\
C^{G\kappa^\mathrm{I}}_{\ell,ij} = \int \f{d\chi}{\chi_\mathrm{m}^2(\chi)}\;w_{i}(\chi)n_{j}(\chi)P_{\delta\kappa^\mathrm{I}}(k; \chi), \nn \\
C^{\kappa^\mathrm{I}\gamma^\mathrm{I}}_{\ell,ij} = \int \f{d\chi}{\chi_\mathrm{m}^2(\chi)}\;n_{i}(\chi)n_{j}(\chi)P_{\kappa^\mathrm{I}\gamma^\mathrm{I}}(k; \chi),
\end{align}
where again the quantities $P_{\kappa^\mathrm{I}}(k; \chi)$, $P_{\delta\kappa^\mathrm{I}}(k; \chi)$ and $P_{\kappa^\mathrm{I}\gamma^\mathrm{I}}(k; \chi)$ could be predicted by some model for intrinsic size/magnitude-density correlations.
 
It is currently unclear if or to what extent these effects will impact cosmic magnification studies. For late-type galaxies at low redshift, a number of studies suggest that the mean size at fixed stellar mass is $\sim 10\%$ higher in the field compared to cluster environments \citep{Maltby2010, FernandezLorenzo:2013ez, Cebrian:2014uc}. It is not clear whether this will persist at higher redshifts, and \citet{Lani:2013hx} find no evidence for environmental dependence of average size at fixed stellar mass for late-type galaxies at redshift $z\sim1$. There is some indication from N-body simulations that the dependence of size on environment may vary with redshift, changing sign for higher redshift galaxies \citep{Maulbetsch:2007ih}. As for the dependence of the luminosity-function on local density for late-type galaxies, there are some reports of a brightening of $M_*$ by $\sim0.5\mathrm{mag}$ from low- to high-density environments (with the other luminosity-function parameters being largely independent of density) \citep{Mo:2004jb, Zucca:2009ic, Croton:2005cl}, whilst others \citep[e.g.,][]{Tempel:2011bd} find that the luminosity function for late-type galaxies is independent of environment.

For early-type galaxies at low redshift $(z < 0.5)$, \citet{Maltby2010} and \citet{HuertasCompany:2012gf} find no evidence for environmental dependence of size at fixed stellar mass for early-type galaxies, whilst \citet{Poggianti:2012kh} find that ellipticals are more compact in denser environments. At higher redshift $(0.5 < z < 2)$, some studies \citep{Cooper2012, Papovich2012, Lani:2013hx, Delaye:2013vf} find that ellipticals are larger in groups or clusters than in the field, whilst others \citep{HuertasCompany:2012gf, Raichoor2011} find no dependence on size with local density. A number of studies suggest that early-type galaxies are brighter in denser environments \citep{Mo:2004jb, Zucca:2009ic, Croton:2005cl}, again with $M_*$ brightening by $\sim0.5\mathrm{mag}$ from the least dense to the most dense environments.

From this somewhat confused picture, it is difficult to build an empirically-motivated model of size-density or magnitude-density correlations or even determine whether they might introduce a non-negligible systematic effect. Nonetheless, we would like to explore to what extent the presence of ISCs (complete with nuisance parameters, as for IAs) may degrade the potential statistical gains from combining magnification with cosmic shear. To this end, we construct a crude model for intrinsic size correlations and suggest an order-of-magnitude upper bound on the amplitude from the current literature. We assume that the mean size of galaxies varies proportionately to local linear matter over-density, and assume an overall variation in the mean of $\sim10\%$ from the densest to the least dense environments (over a range of $-1$ to $+1$ in the linear density contrast) to fix the amplitude. Since we will introduce flexibility into the model and marginalize over a large number of nuisance parameters, we argue that the forecast gains should be relatively insensitive to the fiducial model (notwithstanding the amplitude of the intrinsic signal). However, we stress that this model is not firmly founded in theory or observation, and flag the development of better observations of size-density and magnitude-density correlations in a lensing context as well as better modelling as important future work.

Under these assumptions (and combined with the LA model for IAs), the required intrinsic contributions to the power spectra are:
\begin{align}
&P_{\kappa^\mathrm{I}\kappa^\mathrm{I}}(k;\chi) = \beta^2P^\mathrm{lin}_{\delta\delta}(k; \chi), \nn \\
&P_{\delta\kappa^\mathrm{I}}(k;\chi) = \beta \sqrt{P_{\delta\delta}(k; \chi)P^\mathrm{lin}_{\delta\delta}(k; \chi)}, \nn \\
&P_{\kappa^\mathrm{I}\gamma^\mathrm{I}}(k;\chi) = \beta(-C_\mathrm{I}\rho_0)\sqrt{P^\mathrm{lin}_{\delta\delta}(k; \chi)P^\mathrm{lin}_{\delta\delta}(k; \chi = 0)}.
\end{align}
For the fiducial value of the ISC amplitude we take $\beta = 0.05$. As for IAs, we introduce nuisance parameter grids of the form \eqref{nuisance_grid} into the new intrinsic contributions to the angular power spectra. In summary, the full set of contributions to the estimated shear and convergence power spectra are given by:
\begin{align}
\hat{C}^{\kappa\kappa}_{\ell,ij} &= C^\mathrm{GG}_{\ell,ij} + C^\mathrm{G\kappa^\mathrm{I}}_{\ell,ij} + C^\mathrm{G\kappa^\mathrm{I}}_{\ell,ji} + C^\mathrm{\kappa^\mathrm{I}\kappa^\mathrm{I}}_{\ell,ij} + \delta_{ij}\sigma_{\kappa}^2/\bar{n}_i, \nn \\
\hat{C}^{\gamma_1\gamma_1}_{\ell,ij} &= \hat{C}^{\gamma_2\gamma_2}_{\ell,ij} = C^\mathrm{GG}_{\ell,ij} + C^\mathrm{G\gamma^\mathrm{I}}_{\ell,ij} + C^\mathrm{G\gamma^\mathrm{I}}_{\ell,ji} + C^\mathrm{\gamma^\mathrm{I}\gamma^\mathrm{I}}_{\ell,ij} + \delta_{ij}\sigma_{\gamma}^2/\bar{n}_i, \nn \\
\hat{C}^{\gamma_2\gamma_1}_{\ell,ij} &= C^\mathrm{GG}_{\ell,ij} + C^\mathrm{G\gamma^\mathrm{I}}_{\ell,ij} + C^\mathrm{G\gamma^\mathrm{I}}_{\ell,ji} + C^\mathrm{\gamma^\mathrm{I}\gamma^\mathrm{I}}_{\ell,ij},\nn \\
\hat{C}^{\gamma_1\kappa}_{\ell,ij} &= \hat{C}^{\gamma_2\kappa}_{\ell,ij} = C^\mathrm{GG}_{\ell,ij} + C^\mathrm{G\kappa^\mathrm{I}}_{\ell,ij} + C^\mathrm{G\gamma^\mathrm{I}}_{\ell,ji} + C^\mathrm{\gamma^\mathrm{I}\kappa^\mathrm{I}}_{\ell,ij},
\end{align}
where the intrinsic contributions to the angular power spectra (including nuisance parameter grids) are
\begin{align}
\label{full_angular_power_spectra}
&C^{\gamma^\mathrm{I}\gamma^\mathrm{I}}_{\ell,ij} = \int \f{d\chi}{\chi_\mathrm{m}(\chi)^2}\;n_{i}(\chi)n_{j}(\chi)b_{\gamma^\mathrm{I}}^2(k,z)P_{\gamma^\mathrm{I}\gamma^\mathrm{I}}(k;\chi) \nn \\
&C^{G\gamma^\mathrm{I}}_{\ell,ij} = \int \f{d\chi}{\chi_\mathrm{m}(\chi)^2}\;w_{i}(\chi)n_{j}(\chi)r_{\gamma^\mathrm{I}}(k,z)b_{\gamma^\mathrm{I}}(k,z)P_{\delta\gamma^\mathrm{I}}(k;\chi)\nn \\
&C^{\kappa^\mathrm{I}\kappa^\mathrm{I}}_{\ell,ij} = \int \f{d\chi}{\chi_\mathrm{m}(\chi)^2}\;n_{i}(\chi)n_{j}(\chi)b_{\kappa^\mathrm{I}}^2(k,z)P_{\kappa^\mathrm{I}\kappa^\mathrm{I}}(k;\chi) \nn \\
&C^{G\kappa^\mathrm{I}}_{\ell,ij} = \int \f{d\chi}{\chi_\mathrm{m}(\chi)^2}\;w_{i}(\chi)n_{j}(\chi)r_{\kappa^\mathrm{I}}(k,z)b_{\kappa^\mathrm{I}}(k,z) P_{\delta\kappa^\mathrm{I}}(k;\chi) \nn \\
&C^{\gamma^\mathrm{I}\kappa^\mathrm{I}}_{\ell,ij} = \int \f{d\chi}{\chi_\mathrm{m}(\chi)^2}\;w_{i}(\chi)n_{j}(\chi)r_{\gamma^\mathrm{I}\kappa^\mathrm{I}}(k,z)b_{\kappa^\mathrm{I}}(k,z)b_{\gamma^\mathrm{I}}(k,z) \nn \\
&\;\;\;\;\;\;\;\;\;\;\;\;\;\;\;\;\;\;\;\;\;\;\;\;\;\;\;\;\;\;\;\;\;\;\;\;\;\;\;\;\;\;\;\;\;\;\;\;\;\;\;\;\;\;\;\;\;\;\;\;\;\;\;\;\;\times P_{\kappa^\mathrm{I}\gamma^\mathrm{I}}(k;\chi).
\end{align}
\subsection{Fisher matrix for joint cosmic shear-magnification analysis including physical systematics}
In HAJ13 we derived the Fisher matrix for both a shear only and joint shear-magnification analysis for the purpose of comparison. Here we extend this result to include IAs and ISCs. Following HAJ13 we take our data vectors to contain entries for the estimated harmonic expansion coefficients of the shear and convergence fields (in each tomographic bin) and their complex conjugates,
i.e., $\mathbf{d}^{(\kappa,\gamma)} = (\mathbf{z}^{(\kappa,\gamma)}, \mathbf{z}^{(\kappa,\gamma)*})^\mathrm{T}$
for the combined shear-convergence data and $\mathbf{d}^{(\gamma)} = (\mathbf{z}^{(\gamma)}, \mathbf{z}^{(\gamma)*})^\mathrm{T}$
for the shear only case, where $\mathbf{z}^{(\kappa,\gamma)} = (\dots\hat{\kappa}_{\ell m (i)},\hat{\gamma}_{1,\;\ell m(i)},\hat{\gamma}_{2,\;\ell m(i)}\dots)^\mathrm{T}$ and
$\mathbf{z}^{(\gamma)} = (\dots\hat{\gamma}_{1,\;\ell m(i)},\hat{\gamma}_{2,\;\ell m(i)}\dots)^\mathrm{T}$ contain the full set of relevant harmonic coefficients. The full covariance matrix $\boldsymbol{\Gamma}$ of the data is defined as:
\begin{align}
\boldsymbol{\Gamma} = \langle \mathbf{d} \mathbf{d}^\dagger\rangle &= \left( \begin{array}{cc}
\langle \mathbf{z}\mathbf{z}^\dagger \rangle & \langle \mathbf{z}\mathbf{z} \rangle \\
\langle \mathbf{z}\mathbf{z} \rangle^* & \langle \mathbf{z}\mathbf{z}^\dagger \rangle^*
\end{array}\right) \nn \\
& = \left( \begin{array}{cc}
\mathbf{C} & 0 \\
0 & \mathbf{C}
\end{array}\right),
\end{align}
where in the second line we have used the fact that $\langle \mathbf{z}\mathbf{z} \rangle =0$ and $\mathbf{C} = \langle \mathbf{z}\mathbf{z}^\dagger \rangle\in\mathbb{R}$.
Since different $\ell$ and 
$m$ modes are un-correlated for an all-sky survey, $\mathbf{C}$ will be block diagonal with each $(\ell,m)$-mode contributing one diagonal block (and each $m$-block is identical for fixed $\ell$). Following HAJ13 again and appealing to equation \eqref{full_angular_power_spectra}, we obtain the contribution to $\mathbf{C}$ from a single $(\ell,m)$-mode for the shear-only data vector including IAs:
\begin{align}
\mathbf{C}^{(\gamma)}_\ell = \mathbf{P}^{\gamma\gamma}_\ell\otimes \mathbf{X}^{\gamma\gamma} + \mathbf{\bar{N}}^{-1}\otimes\mathbf{N}_\sigma,
\end{align}
where
\begin{align}
&\mathbf{\bar{N}} = \mathrm{diag}(\bar{n}_1,\;\bar{n}_2\dots),\nn \\
&\mathbf{P}^{\gamma\gamma}_{\ell,ij} = C^\mathrm{GG}_{\ell,ij} + C^\mathrm{G\gamma^\mathrm{I}}_{\ell,ij} + C^\mathrm{G\gamma^\mathrm{I}}_{\ell,ji} + C^\mathrm{\gamma^\mathrm{I}\gamma^\mathrm{I}}_{\ell,ij}, \nn \\
&\mathbf{N}_\sigma = \mathrm{diag}\left(\sigma_\gamma^2, \sigma_\gamma^2\right), \nn \\
&\mathbf{X}^{\gamma\gamma} = \left( \begin{array}{cc}
1 & 1 \\
1 & 1 \end{array} \right),
\end{align}
and $\otimes$ is the tensor product. For the magnification-shear data vector, this is extended to:
\begin{align}
\mathbf{C}^{(\kappa\gamma)}_\ell = \mathbf{P}^{\gamma\gamma}_\ell\otimes \mathbf{X}^{\gamma\gamma}+ \mathbf{P}^{\kappa\gamma}_\ell\otimes \mathbf{X}^{\kappa\gamma} + &\mathbf{P}^{\gamma\kappa}_\ell\otimes \mathbf{X}^{\gamma\kappa}  + \mathbf{P}^{\kappa\kappa}_\ell\otimes \mathbf{X}^{\kappa\kappa}\nn \\
& + \mathbf{\bar{N}}^{-1}\otimes\mathbf{N}_\sigma,
\end{align}
where
\begin{align}
&\mathbf{\bar{N}} = \mathrm{diag}(\bar{n}_1,\;\bar{n}_2\dots),\nn \\
&\mathbf{P}^{\gamma\gamma}_{\ell,ij} = C^\mathrm{GG}_{\ell,ij} + C^\mathrm{G\gamma^\mathrm{I}}_{\ell,ij} + C^\mathrm{G\gamma^\mathrm{I}}_{\ell,ji} + C^\mathrm{\gamma^\mathrm{I}\gamma^\mathrm{I}}_{\ell,ij}, \nn \\
&\mathbf{P}^{\kappa\gamma}_{\ell,ij} = C^\mathrm{GG}_{\ell,ji} + C^\mathrm{G\kappa^\mathrm{I}}_{\ell,ji} + C^\mathrm{G\gamma^\mathrm{I}}_{\ell,ij} + C^\mathrm{\gamma^\mathrm{I}\kappa^\mathrm{I}}_{\ell,ji}\nn \\
&\mathbf{P}^{\gamma\kappa}_{\ell,ij} = C^\mathrm{GG}_{\ell,ij} + C^\mathrm{G\kappa^\mathrm{I}}_{\ell,ij} + C^\mathrm{G\gamma^\mathrm{I}}_{\ell,ji} + C^\mathrm{\gamma^\mathrm{I}\kappa^\mathrm{I}}_{\ell,ij}\nn \\
&\mathbf{P}^{\kappa\kappa}_{\ell,ij} = C^\mathrm{GG}_{\ell,ij} + C^\mathrm{G\kappa^\mathrm{I}}_{\ell,ij} + C^\mathrm{G\kappa^\mathrm{I}}_{\ell,ji} + C^\mathrm{\kappa^\mathrm{I}\kappa^\mathrm{I}}_{\ell,ij}\nn \\
&\mathbf{N}_\sigma = \mathrm{diag}\left(\sigma_\kappa^2, \sigma_\gamma^2, \sigma_\gamma^2\right), \nn \\
&\mathbf{X}^{\gamma\gamma} = \left( \begin{array}{ccc}
0 & 0 & 0\\
0 & 1 & 1\\
0 & 1 & 1\end{array} \right),\;\;\mathbf{X}^{\kappa\gamma} = \left( \begin{array}{ccc}
0 & 1 & 1\\
0 & 0 & 0\\
0 & 0 & 0\end{array} \right),\nn \\
&\mathbf{X}^{\gamma\kappa} = \left( \begin{array}{ccc}
0 & 0 & 0\\
1 & 0 & 0\\
1 & 0 & 0\end{array} \right),\;\;\mathbf{X}^{\kappa\kappa} = \left( \begin{array}{ccc}
1 & 0 & 0\\
0 & 0 & 0\\
0 & 0 & 0\end{array} \right).
\end{align}
Recall the Fisher matrix is the expectation of the second derivative of the negative log-likelihood with respect to the model parameters. Here we assume the data vector to be Gaussian distributed with fixed means,
so the Fisher matrix can be straightforwardly computed from the covariance matrix and its derivatives \citep{Tegmark1997}:
\begin{align}
\mathbf{F}_{\alpha\beta} &= \f{1}{2}\mathrm{Trace}\left[\mathbf{\boldsymbol{\Gamma}}^{-1}\boldsymbol{\Gamma}_{,\;\alpha}\boldsymbol{\Gamma}^{-1}\boldsymbol{\Gamma}_{,\;\beta}\right], \nn \\
&= \mathrm{Trace}\left[\mathbf{C}^{-1}\mathbf{C}_{,\;\alpha}\mathbf{C}^{-1}\mathbf{C}_{,\;\beta}\right]\;.
\end{align}\label{Fisher1}
Exploiting the block-diagonal property of $\mathbf{C}$, the Fisher matrix can be written as a sum over modes:
\begin{align}
\mathbf{F}_{\alpha\beta} = f_\mathrm{sky} \sum_{\ell_\mathrm{min}}^{\ell_\mathrm{max}}\left(\ell + \f{1}{2}\right)\left[ \mathbf{C}_\ell^{-1}\mathbf{C}_{\ell,\;\alpha}\mathbf{C}_\ell^{-1}\mathbf{C}_{\ell,\;\beta}\right],
\end{align}
where we have also included a factor $f_\mathrm{sky}$ to approximately account for incomplete sky coverage.
\subsection{Fisher matrix forecasts}
We consider a $\num[group-separator={,}]{15000}$ square degree survey similar to that proposed for the ESA Euclid mission. We assume a redshift distribution $n(z) \propto z^2 \mathrm{exp}[-(1.41z/z_m)^{1.5}]$, with a median redshift $z_m = 0.9$ and a mean number density $\bar{n}  = 30$ per square arcminute. We take Gaussian photometric redshift uncertainties, with redshift-dependent dispersion $\sigma_z = 0.05(1+z)$. We use CAMB to compute the matter power spectrum, and vary the following cosmological parameters: $\Omega_b, \Omega_c, \Omega_\Lambda, h, w_0, w_a, n_s, 10^9A$, being, respectively, the density parameters in baryons, Cold Dark Matter and Dark Energy, the Hubble parameter in units of $100\mathrm{km s^{-1} Mpc^{-1}}$, the Dark Energy equation of state parameters $(p/\rho = w_0 +w_a(1-a)$, where $a$ is the scale factor), the scalar spectral index, and the amplitude of fluctuations. We assume a dispersion in $\kappa$ estimation of $\sigma_\kappa = 0.8$; this is the weighted average of the Fisher forecast $\sigma_\kappa$ from the fitted size-magnitude distributions (with smooth selection function) and redshift bins described in \S \ref{joint_dist}-\ref{dispersion}, for both early- and late-types. For ellipticity we take $\sigma_e = 0.38$ for the intrinsic (complex) ellipticity dispersion (estimated from the data). We consider a tomographic set-up with 10 bins between redshifts 0 and 2, with equal numbers per bin, and $\ell$-modes upto $\ell = 3000$.

To explore the impact of systematics (IAs and ISCs) we consider three scenarios: (1) no systematics in either shear or magnification, (2) marginalizing over nuisance parameters for IAs only, assuming no systematics for the magnification signal, and (3) marginalizing over both IA and ISC nuisance parameters. When marginalizing over nuisance parameters we consider $5\times 5$ grids in $(k,z)$ plus amplitude parameters, as described in \S \ref{sec:intrinsic_alignments}-\ref{sec:isc}, with broad (uninformative) Gaussian priors on all nuisance parameters. This accounts for the substantial uncertainty in our current knowledge of IAs and ISCs. Future observations may place priors on these nuisance parameters. To estimate the relative statistical power in each case we compare the Figure-of-Merit (FoM) for Dark Energy, defined to be the inverse of the (area/$\pi$) of the $1\sigma$ contours of the expected likelihood in the $w_0,w_a$ plane, marginalised over all other parameters. The results are summarized in Table \ref{tab:fom_values}.
\begin{table}
\caption{Dark Energy Figure of Merit (FoM) values computed from the Fisher Matrices for various scenarios regarding systematics.}
\begin{tabular}{cccc}
\hline
Scenario: & No Systematics & IA & IA and ISC \\
\hline
shear & 199 & 46 & -- \\
magnification & 86 & -- & -- \\
shear+magnification & 224 & 114 &  76 \\
\hline
\end{tabular}
\label{tab:fom_values}
\end{table}

In scenario (1) (no systematics) we find that magnification should have roughly $50\%$ of the signal-to-noise compared to shear (computed as the ratio of the square root of the FoM) and the FoM for a combined shear-magnification analysis is $13\%$ larger compared to shear only. Shear is evidently statistically dominant due to the relatively large dispersion on $\kappa$ estimation, $\sigma_\kappa=0.8$ compared to $\sigma_e=0.38$ for (complex) shear.

In scenario (2), marginalizing over IA nuisance parameters, the anticipated improvement in considerably larger. For shear only, marginalizing over the 52 nuisance parameters (two $5\times 5$ grids plus two amplitude parameters) degrades the FoM by a factor of $\sim 4$ (i.e., the forecast error bars on cosmological parameters are increased by a factor $\sim 2$). For shear+magnification, marginalizing over IAs degrades the FoM by a more modest factor of $2$; in this set-up, magnification both adds additional information and helps to calibrate the IAs, reducing the impact of nuisance parameter marginalization. With no systematics on magnification but accounting for IAs, we find that magnification could out-perform shear and adding magnification information to shear improves the FoM by a factor of $2.5$. This is the best-case-scenario for magnification. Note that this result is strongly dependent on the number of nuisance parameters marginalized over; for two $2\times 2$ grids plus amplitudes (10 nuisance parameters) the gain from adding magnification to shear+IAs is reduced to a more modest $30\%$.

\begin{figure}
\includegraphics[width = 8.22cm]{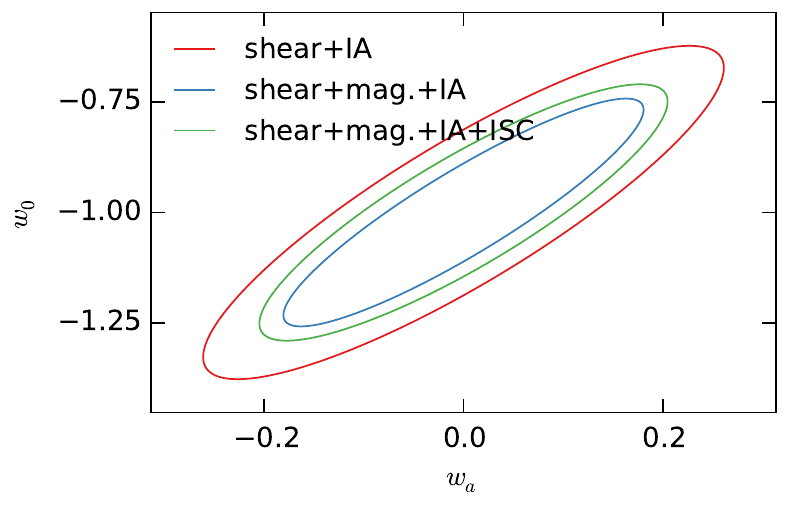}
\caption{Fisher forecast $1\sigma$ error ellipses on the Dark Energy parameters, marginalized over all other parameters, for various scenarios regarding systematics: shear with IAs (red), combining magnification without systematics and shear with IAs (blue), and combining magnification with ISCs and shear with IAs (green). In all cases systematics are parameterized with $5\times 5$ grids in scale and redshift plus amplitude nuisance parameters; this amounts to 52 nuisance parameters for IAs, and 104 nuisance parameters for IA and ISC together.}
\label{fig:blob}
\end{figure}

In scenario (3) we marginalize over nuisance parameters for both IAs and ISCs. The addition of magnification information (without systematics) to shear+IAs improves the FoM by a factor of 2.5. Whilst adding in ISCs degrades this gain, a substantial fraction of the gain is preserved, due in part to the presence of different systematics on shear and magnification helping to calibrate each other and hence reducing the impact of nuisance parameter marginalization. The FoM for magnification+shear+ISC+IA is enhanced by roughly $65\%$ compared to shear+IA. It is important to point out this improvement is strongly dependent on the amplitude of the ISC terms. Reducing the amplitude of the ISC by a factor of $10$ attenuates the gain to $\sim25\%$. This is due to the competition between nuisance parameter marginalization degrading the FoM and cosmological dependence of the intrinsic terms adding extra signal; as the amplitude of the intrinsic terms is reduced, the effect of nuisance parameter marginalization becomes relatively more dominant. Reducing the amplitude beyond a factor of $10$ does not degrade the gain further, so we anticipate between $25\%$ and $65\%$ improvement in the FoM from adding magnification when accounting for both IAs and ISCs, for the set-up with $5\times 5$ nuisance parameter grids in scale and redshift plus amplitude parameters. These results are shown in Fig. \ref{fig:blob}.

Whilst the realistic scenario of a shear+magnification analysis including both IAs and ISCs still performs worse than a naive shear-only analysis which ignores IAs, it still significantly out-performs shear+IA by a factor of upto 1.65. Like in a shear-only analysis, there is great potential for further `self-calibration' of systematics in a shear+magnification analysis by using the large-scale structure (LSS) information that comes for free in any weak lensing survey. For cosmic shear, the self-calibration from combining shear with LSS could potentially restore all of the information lost by marginalizing over IAs \citep{Joachimi:2009id}. The impact of the same process on a shear+magnification analysis would be an interesting study.

When marginalising over IAs and ISCs we used $5\times 5$ grids in scale and redshift giving the models a significant degree of flexibility; the amount of freedom given to the models for IAs and ISCs deserves some discussion. Great care must be taken to ensure that the models for IAs and ISCs have sufficient flexibility so that we do not gain cosmological information from the IA and ISC power spectra, since we do not have robust physically motivated models that we can rely on for cosmological inference. With the amplitude of the ISC power spectrum taken here (see \S \ref{sec:isc}), we find that reducing the number of nuisance parameters below $5\times 5$ grids in scale and redshift moves us quickly into the regime where we are gaining information from the systematic contributions to the power spectra (i.e. the inclusion of ISCs could increase the FoM). This is not realistic based on our current ignorance of the IA, and particularly the ISC, signal. To avoid unrealistically precise constraints we must employ the flexible nuisance parameter grids used in this work. Current detections of IAs for early-type galaxies have at least $\sim10\%$ uncertainties \citep{Singh2014}, and IA measurements for the statistically more dominant late-types, whilst consistent with zero, come with considerable uncertainty (see e.g. \citet{Hirata2007, Mandelbaum2011}). Dedicated IA measurements have also only been made at low redshift; assuming that these observations can be extrapolated to high redshift is not justified, motivating multiple nuisance parameters in redshift. Furthermore, even with the relatively large uncertainties in current IA measurements, it’s clear that the Linear Alignment model (used in this work) \citep{Hirata2004} and Non-Linear Alignment model \citep{Bridle2007} fail on small scales \citep{Schneider2010, Singh2014, Sifon2015}, motivating multiple nuisance parameters to give the models flexible scale-dependence. There is also evidence for luminosity dependence in current intrinsic alignment measurements \citep{Hirata2007, Joachimi2011, Singh2014}, which is not accounted for in our baseline model, motivating further flexibility. Even less is known about ISCs than IAs, so a flexible nuisance parameter grid is clearly required for ISCs. This motivates the $5\times 5$ nuisance parameter grids in scale and redshift used in this work.
 
\section{Conclusions}
We have derived the posterior distribution for the convergence field from a measured galaxy size, magnitude and redshift. In general, we showed that this requires detailed knowledge of both the intrinsic size-magnitude distribution as a function of redshift and the selection function in the size-magnitude plane. The width of this posterior distribution (i.e. how well we can recover $\kappa$ from sizes and magnitudes) depends critically on the shape of the size-magnitude distribution. For comparison with the Bayesian approach, we developed a simple unbiased estimator for convergence by taking an optimally weighted linear combination of (mean subtracted) size and magnitude, where optimal weights can be calibrated directly from data.

By building a simple model for the size-magnitude distribution and fitting this model to the CFHTLenS galaxy sample, we find that the convergence field should be recoverable with a typical dispersion of $\sim 0.8$ (for a single source). Compared to the two-component shear, which can be estimated with a typical dispersion of $\sim0.38$ (due to the intrinsic scatter in galaxy ellipticities), it is clear that shear will be statistically more powerful than magnification. Indeed, we find that in the absence of systematics (IAs or ISCs) magnification should have $\sim50\%$ of the signal-to-noise compared to cosmic shear, and combining shear and magnification improves the Dark Energy Figure-of-Merit by $\sim13\%$ over shear-only (at the Fisher matrix level).

It's possible that magnification using sizes and magnitudes will be subject to physical systematic effects - size-density and/or magnitude-density correlations (ISCs) - not dissimilar to intrinsic alignments (IAs) in the case of cosmic shear. However, it's not yet clear whether or to what extent such effects will be a limitation for cosmic magnification. To study the possible impact of systematics on the relative statistical power of shear versus magnification, we proposed a crude model for ISCs and take the linear alignment model for IAs, both parameterized with grids of free nuisance parameters. In the case where IAs are present but ISCs are not, we find that magnification has larger signal-to-noise compared to shear, and combining magnification with shear could improve the FoM by up to a factor of $2.5$. In the scenario where we marginalize over nuisance parameters for both IAs and ISCs, combining magnification and shear information still has the potential to give substantial improvements in the FoM over shear only and we anticipate a gain of between $25\%$ and $65\%$ depending on the amplitude of the ISC power spectrum.

\section*{acknowledgments}
We thank Hendrik Hildebrandt, Malte Tewes, Lance Miller, Fabian Schmidt, Till Hoffmann, Omar Almaini and Meghan Gray for useful discussions.

\bibliographystyle{mn2e_williams}
\bibliography{sizemag}

\end{document}